\documentclass[12pt,a4paper]{article}

\usepackage{etex}
\usepackage[T1]{fontenc}
\usepackage[german,english]{babel}
\usepackage[ansinew,utf8]{inputenc}
\usepackage[final]{graphicx}
\usepackage{mathptmx}
\usepackage{amsmath}
\usepackage{amssymb}
\usepackage{amsfonts}
\usepackage{accents}
\usepackage[final]{graphicx}
\usepackage{booktabs}
\usepackage{rotating}
\usepackage[round]{natbib}
\usepackage{color}
\usepackage{eepic}
\usepackage{epic}
\usepackage{multirow}
\usepackage{bm}
\usepackage[titletoc,title]{appendix}
\usepackage{bbm}
\usepackage{url}
\usepackage[flushleft]{threeparttable}
\usepackage{longtable}
\usepackage{setspace}
\usepackage{hyperref}
\usepackage{subcaption}
\usepackage{caption}
\usepackage{dcolumn}
\usepackage{setspace} 
\newcommand\fnote[1]{\captionsetup{font=scriptsize}\caption*{#1}}

\usepackage[top=1.62in, bottom=1.62in, left=1.2in, right=1.2in]{geometry}

\usepackage{siunitx}
\usepackage{textcomp}
\usepackage{microtype}

\begin{document}

\title{Dealing with Uncertainty: The Value of Reputation in the Absence of Legal Institutions\thanks{We are grateful to Berno Buechel, Stefan Buehler, Maximilian Conze, Philemon Kraehenmann, Frank Pisch, Alexander Rasch, and seminar participants at EARIE 2018 (Athens), IIOC 2019 (Boston), Workshop on the Economics of Digitization 2019 (Louvain-la-Neuve), Jornadas de Economia Industrial 2019 (Madrid), Swiss IO Day 2019 (Bern), at the Duesseldorf Institute for Competition Economics (DICE), the University of Konstanz, and the University of St.\ Gallen for helpful discussions and comments. The usual disclaimer applies.}}

\author{Nicolas Eschenbaum and Helge Liebert\thanks{Nicolas Eschenbaum:\ University of St.\thinspace Gallen, Institute of Economics (FGN), Varnbüelstrasse~19, 9000 St.\thinspace Gallen, Switzerland (nicolas.eschenbaum@unisg.ch). Helge Liebert:\ University of Zürich, Department of Economics, Schönberggasse 1, 8001 Zürich, Switzerland (helge.liebert@uzh.ch).} \\ \phantom{x}}

\vspace{0.4cm}

\date{\normalsize May 2021}

\maketitle

\begin{abstract}
  \noindent This paper studies reputation in the online market for illegal drugs in which no legal institutions exist to alleviate uncertainty. Trade takes place on platforms that offer rating systems for sellers, thereby providing an observable measure of reputation. The analysis exploits the fact that one of the two dominant platforms unexpectedly disappeared. Re-entering sellers reset their rating. The results show that on average prices decreased by up to 9\% and that a 1\% increase in rating causes a price increase of 1\%. Ratings and prices recover after about three months. We calculate that identified good types earn 1,650 USD more per week.
\end{abstract}

\textbf{Keywords:} Reputation, institutions, uncertainty, dark web, drugs

\emph{JEL-Classification:} L14, L15, L81, K42

\addtocounter{page}{-1}
\thispagestyle{empty}
\newpage
\thispagestyle{plain}

\doublespacing

\section{Introduction}
\label{sec:Introduction}

For individuals to engage in mutually beneficial exchange, they must trust the counterparty to fulfill its promised role in the transaction. One way to establish trust is to rely on the reputation of an individual as an honest trader. By making transactions conditional on the counterparties' reputation, past actions are linked to future trades. If the value of reputation for future trades is sufficiently high, short-run incentives to cheat can be overcome. 

The use of a reputational mechanism to diminish uncertainty in trade has been shown to be an important factor in a variety of settings \cite[e.g.,][]{milgrom1990,greif1989}, and previous work has demonstrated that reputation can overcome uncertainty and allow markets to function when formal or legal institutions enforced by a (state) authority are absent \cite[e.g.,][]{greif1994,leeson2007}. However, while the theoretical mechanism is well-understood, empirical evidence on reputation in such unique settings in which legal institutions are absent is scarce. Reputation is inherently hard to measure and in many settings it is difficult to pin down the extent to which reputation substitutes or complements legal institutions. The rise of online trade has allowed reputation to be measured in the form of publicly observable ratings and a large literature has developed that estimates returns to reputation \cite[e.g.,][]{cabral2010}. But the role of the reputational mechanism in overcoming uncertainty is still difficult to disentangle from the effects of legal institutions.

In this paper, we provide empirical evidence on the value of reputation in a market devoid of contracting institutions deriving from a legal system. We make use of a unique dataset of the online market for illegal drugs. Due to the illegal nature of the transactions and strong need for market participants to remain anonymous, contracts are  unenforceable. Trade is conducted on online sales platforms that provide a rating system for merchants in a form that is familiar to any user of e.g. Amazon.
We focus our attention on the two dominant sales platforms from 2014 to 2015, jointly covering more than 90\% of the market. We exploit the fact that one of the two platforms unexpectedly exited the market and track sellers that re-entered the market and began selling on the remaining platform in the aftermath. These sellers were forced to open a new account and hence reset their rating in the process. 

We evaluate the effects of the forced market re-entry and document three main results. First, we estimate an event study specification and find that re-entering the market leads to a decrease in unit prices of up to 9\% on average. The price decrease occurs gradually over the first few weeks, peaks after 7-9 weeks, and disappears after 12-13 weeks. The time it takes for the price effect to vanish is exactly in line with the average time it takes for a new seller to establish a high rating. We calculate that this implies a revenue loss of 4,800 USD on average. Second, we exploit the exogenous variation in rating due to the forced re-entry to estimate the ratings effect on price and find an average elasticity of $1$. This implies that an improvement in rating by one standard deviation causes an increase in prices of 12\%.  We further provide evidence that as the rating recovers, the effect of rating on price begins to disappear. Third, we document that re-entering sellers were representative of highly-reputable sellers, i.e., ``good'' types, prior to the exit and new entrants after, allowing us to obtain an explicit estimate of the value of being a good type in a separating equilibrium instead of a pooling one in a setting with unobservable quality. We calculate that the additional revenue on average is 1,650 USD per week. In addition, our empirical approach implicitly shows that reputation is (partially) non-transferable between online platforms. To the best of our knowledge, this has not been documented explicitly before.

The magnitude of the effects we find highlights the importance of reputation in a market in which contracts are unenforceable and no legal system is available. Similar analysis conducted for legal sales platforms has often documented smaller effects \citep[e.g.,][]{cabral2010,cai2014,resnick2002}. The reputational cost of the forced re-entry for sellers can also be interpreted as the cost of being ostracized. Ostracism has been documented to be an important tool to enforce common rules for trade and honest behavior \citep{milgrom1990,greif1989,north1991,benson1989}. However, ostracism as an enforcement mechanism is unlikely to be effective in our setting due to the degree of anonymity; voluntarily re-entering the market under a new pseudonym is always feasible. This suggests that our findings likely downplay the cost of being ostracized in markets in which sellers are not anonymous.

To conduct our analysis, we make use of webscrapes of individual offers on the two most popular platforms for illegal merchandise during 2014 and 2015: ``Agora'' and ``Evolution.'' We further add data based on API requests of the darknet search engine ``Grams.'' The resulting dataset provides a full overview of the supply of drugs on the two dominant darknet trading platforms and covers the time period of June 2014 until July 2015. It contains information on the prices and quantities of each offer, the type of drug sold, whether the offer allows use of escrow services, the country the good is shipped from, as well as the seller's rating, size, name and public PGP (``pretty good privacy'') key for encrypted communication.


In mid-March 2015, the administrators of Evolution executed what is known as an ``exit-scam'' and absconded with an estimated \$34 million in bitcoins (at the time) stolen from their traders. That is, they unexpectedly shut down access of buyers and sellers to their respective bitcoin wallets on the platform and took the website offline. Vendors selling exclusively on the Evolution platform were subsequently forced to migrate to  Agora or exit the market altogether. We exploit our knowledge of sellers' public PGP keys to link vendor accounts over time and across platforms. This allows us to track sellers that sold on Evolution prior to the exit and migrated to Agora following the exit.
We rely on the exogenous variation induced by the platform exit to study the impact of re-entry and rating on price.

Our work in this paper makes three contributions to the literature. One, we study a unique black market that has received limited attention so far in economics, in which legal institutions are replaced by a reputational mechanism. Our estimates provide empirical evidence on the role reputation plays in overcoming a lack of contracting institutions. This possibility has been emphasized before in, among others, \citet{acemoglu2005},\citet{greif1989},\citet{macleod-2007}, \citet{north1991}, and explicitly documented in a variety of unique settings, such as emerging markets \citep{gao2017}, medieval merchant guilds \citep{greif1994}, a private code of law for merchants in the middle ages \citep{milgrom1990}, or pirate organizations \citep{leeson2007}. The existence of the observable ratings system allows us to quantify the effects. To the best of our knowledge, we are the first to investigate reputation empirically in a setting devoid of legal institutions and provide causal estimates.

Only few authors have previously studied the online market for illegal drugs in economics.\footnote{However, there exists a larger literature in economics on the trade of illegal drugs offline, as well as on the effects of drug liberalization policies \citep[e.g.][]{galenianos2017,jacobi2016,adda2014}. There is also a small literature in computer science and criminology on darknet marketplaces \citep[e.g.][]{soska2015,aldridge2014,barratt2016}.} \citet{bhaskar2017} document the evolution of the online drug trade and the darknet platforms over time. \citet{janetos2017} analyze the Agora platform and develop a structural model of reputation-formation. In contrast, we do not focus on the mechanics of reputation-building by sellers, but on the ability of the reputational mechanism in overcoming the uncertainty buyers face and the corresponding value of reputation for sellers. \citet{espinosa-2018} investigates scamming of buyers on a smaller darknet platform. Both \citet{espinosa-2018} and \citet{bhaskar2017} also document estimates of reputational effects using fixed-effects regressions that are substantially smaller than our findings and which are broadly in line with results documented for legal markets. We investigate this difference to our findings and provide estimates of basic fixed effects regressions, also finding no or small effects of rating on price only.

Two, we provide reduced-form estimates of reputation effects on online-sales platforms using a novel approach that exploits the reputational shock experienced by sellers forced to re-enter the market following the Evolution exit. A sizable literature has developed that estimates the effects of reputation for sellers in online markets (see \citet{tadelis2016} and \citet{cabral2012} for surveys). Much of the literature documents that a high rating or a large number of positive versus negative feedback is generally associated with higher sales rates  and completion rates for transactions \citep[e.g.][]{cabral2010,cai2014}, but that the effect on price is relatively small \citep[e.g.][]{resnick2002,cabral2010,houser-wooders-06}. However, some studies are exceptions to this pattern and document substantial reputational premia that are more in line with our own estimates \citep[e.g.][]{jolivet2016, fan2016}. \cite{fan2016} document a substantial effect of rating on revenue for established sellers on Taobao, but a negative effect for new sellers. Their work suggests that new sellers respond to improving rating by cutting prices to increase transaction volume. In our analysis, we find that re-entering sellers with lower ratings set lower prices and do not observe a corresponding change in the rate of sales. Lastly, we make use of a novel approach to estimate the ratings effect on price based on multi-platform data and exogenous variation in platform availability. Similar to our analysis, \cite{benson2020} exploit the disappearance of reputation by studying instances when Turkopticon servers crash and show that only employers with good reputation are affected. We document that it is highly-reputable sellers that re-enter the market and suffer the consequence of the sudden loss in reputation.

Three, our estimates shed light on the value of playing a separating instead of a pooling equilibrium in a setting with unobservable quality. Previous work that documents signaling in online markets generally relies on mechanisms that are not available in the market we study (e.g.\ reward-for-feedback \citep{tadelis2020} or reserve prices \citep{kawai2014}).
We observe in our data that (i) identified low-reputation sellers tend to exit the market quickly, while identified high-reputation sellers tend to stay in the market long, and (ii) tracing back these two sets of sellers to their early days in the market does not reveal consistent price differences between ``good'' and ``bad'' types. In addition, the theoretical literature on quality signaling establishes that an important condition for signaling to occur in equilibrium is that low-quality sellers must prefer such a full-information equilibrium to a pooling one \citep[e.g.][]{daughety2008b,janssen2015,milgrom1986}. The fact that sellers that accumulate a low rating leave the market early strongly suggests that this condition is not satisfied in our setting.   We use our estimate of the effect of rating on prices to calculate the implied revenue loss per week for an established, high-quality  seller that is forced to sell at the rating of an entrant in the market.

The remainder of this paper is structured as follows. Section~\ref{sec:Market} discusses the setup of the online drug market. Section~\ref{sec:Descriptives} explains how the data was collected and processed, and establishes stylized facts about the nature of seller ratings and the determination of prices. Section~\ref{sec:Approach} details the empirical approach and discusses the identification of the ratings effect.  Section~\ref{sec:Results} documents the results and sensitivity checks. Finally, section~\ref{sec:Conclusion} concludes.


\section{The market}
\label{sec:Market}
Trade in illegal drugs has moved online in the past decade to a significant degree. \citet{foley-et-al-2019} estimate that illegal trade conducted using bitcoin amounts to \$76 billion per year, while the 2017 Global Drug Survey documents that in the UK in 2017, around a quarter of respondents report purchasing drugs online \citep{barratt2016}. These online drug trades take place on sales platforms located on the dark web. Following a purchase on such a platform, the product is shipped by mail to the customer.\footnote{\citet{soska2015} study the first of these platforms, ``Silk Road'', and estimate that the website at its height in 2013 had an annual revenue of more than \$100 million.}

These darknet platforms are characterized by four features. First, they are located on the Tor network. Tor (``the onion router'') makes use of a private network that directs an internet users signal across different relays and encrypted nodes before reaching the intended destination, making it very difficult to track the site or its users. Second, they enable and encourage their users to communicate using PGP (`pretty good privacy') encryption. Third, transactions can only be conducted using a cryptocurrency (usually bitcoin). Each seller or buyer can deposit and withdraw bitcoins from a wallet attached to their account on the site in order to make payments. Fourth, a centralized feedback and rating system is provided, in which buyers can leave feedback for sellers they have bought from. In addition, platforms often provide an escrow system that sellers can offer to customers. Instead of making the payment directly to the vendor's on-site wallet, the buyer sends his bitcoins to a wallet of the platform. Then, the vendor sends the merchandise and after the buyer confirmed its arrival, the platform transfers payment to the seller.\footnote{If the buyer does not signal the shipment to have concluded, the payment is automatically released after a waiting period of a few weeks. In addition, platforms often offer mediation services in case of disputes (e.g.~Agora). Such an escrow system is not unique to illegal online markets. Airbnb for example holds a buyers payment until 24 hours after check-in ``to make sure everything is as expected'' \citep{airbnb}. See \autoref{fig:escrow} in the appendix for an overview of the escrow system on Silk Road.}  Sellers can also choose to forego this system and require buyers to send the funds directly prior to shipment of the merchandise.

The illegal marketplaces we study strongly encourage buyers and sellers to encrypt their communication. When buyers choose to make a purchase, they must provide the seller with an address for the shipping of the merchandise. Doing so in the clear given the illegal nature of the trade poses an additional risk for buyers. Consequently, vendors are required to  provide their public PGP key for buyers to use in their offerings on the platform, so that each vendor account on a platform is linked to a specific public PGP key. Private and public keys are unique and highly complex. These features of PGP keys provide us with a unique identifier that allows matching vendor accounts across platforms and time. \autoref{fig:pgpkey} in the appendix shows an example of a public PGP key.\footnote{Public-key cryptography relies on two separate keys, one public and one private. Communication is done by encrypting the information sent with the openly published public key of the receiver. Decrypting can then only be done by using the private key, which is kept secret. Public-key cryptography is designed to ensure that re-engineering the private from the public key is computationally infeasible.}





On the surface, darknet platforms are structured in a way familiar to any user of eBay or Amazon (see \autoref{fig:agoscreens} for examples from the ``Agora'' platform). Sellers can open accounts and create product listings. Each listing contains a description of the product on offer and a price set by the seller, as well as information on the shipping origin of the merchandise and the sellers rating and number of previous sales made. Buyers in turn can browse the listings by selecting the relevant category of products, or using the sites' search function.

In contrast to clearnet markets, there is a great deal of darknet platform turnover. At any given point in time there may be dozens of different marketplaces active on the darknet. However, trade is predominantly conducted on a few large platforms. \citet{bhaskar2017} document the lifetime of 88 separate platforms from 2011-2015 and demonstrate that the vast majority of them were (very) small in terms of market size and had a very short lifespan. The few large platforms in turn dominate the market when active and operate for a significantly longer time period of at least a year. Platforms exit for multiple reasons, among others a shutdown by the authorities (e.g.~Silk Road), an exit-scam (e.g.~``Evolution''), or voluntarily due to for example security concerns (e.g.~Agora). During the time frame studied in this paper, the Agora and Evolution platforms were the dominant players in the market. Following the Evolution exit-scam in March 2015, Agora continued to be the largest platform until its voluntary and pre-announced exit in August 2015.

In an exit-scam, the platform administrators steal the currency held by buyers and sellers on their respective platform accounts and take the platform offline. This is possible because traders conduct their business using on-site accounts for which they are granted control by the platform operators. In order to make a purchase, buyers place bitcoins in their platform account to then make transactions. Similarly, sellers have to withdraw their earnings from their platform account. Importantly, even though buyers and sellers have nominal control over their on-site wallets, this is purely at the operators' discretion and access can be revoked at will. Any funds that are stored in an on-site account are ultimately controlled by the platform operators. In mid-March of 2015, Evolution began to disallow withdrawals of bitcoins from wallets and accounts on the platform, citing technical difficulties. Escrow accounts were similarly frozen and inaccessible. Within a week the site went offline. Estimates suggest that the site administrators stole around 130,000 bitcoins from their users, worth at the time approximately \$34 million (and about 7.5 billion in May 2021). The exit was highly unexpected, since Evolution was the largest platform in the market and was known for stability and professionalism. Cursory examination of discussion forums on darknet platforms at the time suggests that it took 2-3 days for traders to start to become aware of the scam occurring.

However, traders quickly migrated to other platforms. At the time, Agora was the only remaining large and dominant marketplace and saw a sudden increase in sellers following the Evolution exit (see \autoref{fig:marketsize}). Sellers forced to re-enter the marketplace had to register a new account and hence lost their reputation in the process.\footnote{Agora and Evolution operated in the exact same way and offered the same services to their users. They also had the same fee structure for operating a seller account. \autoref{tab:platf-diff} in the appendix documents the average price, median quantity, and number of vendors and items prior to the exit on either platform. We do not observe any patterns or significant differences that would indicate variation in how the two platforms operated in the market.} We exploit this ratings reset of sellers who move platforms to estimate the effect of ratings on price and we track sellers across platforms by linking their accounts as described above.



\section{Data and descriptives}
\label{sec:Descriptives}
We make use of daily webscrape data from the two darknet platforms Agora and Evolution, as well as of daily API requests of the darknet search engine Grams (see \autoref{fig:grams} in the appendix for a screenshot of the website). Our full data covers the time period of July 2014 to July 2015. The Grams data allows us to obtain information on the supply of goods on the two platforms. Illegal drugs account for the largest share of merchandise on offer.\footnote{Drugs and electronic goods (such as eBooks or credentials for hacked Netflix accounts) are by far the two largest categories, making up around 99\% of the market.} For each item on offer we observe the title, price, product category, and shipping origin, as well as the vendor's name and public PGP key. We add to this the vendor's rating, number of reviews, a measure of total sales, and the item description from the platform scrapes. The resulting dataset provides a unique overview of the black market for illegal drugs over the course of a year.

Items advertised on the darknet platforms are placed in separate product categories, allowing us to distinguish different types of drugs. We focus on homogeneous goods within each drug type and extract from the item titles and descriptions information on the quantity being sold and the size of the batch. \autoref{fig:evo-lsdoffer} shows an example of an offer for Cocaine. In this instance, we determine that the quantity sold in the offer is $1$ gram.

Between July 2014 and July 2015, we observe eight main product categories of illegal drugs, namely cannabis, MDMA, cocaine, amphetamine, methamphetamine, heroin, LSD, and ketamine, and a total of 37,057 unique offers of drugs made by 3,005 separate vendors. Since the accessibility of the sites varies over time, our dataset sometimes contains gaps between observations of offers. These missing data patterns are unlikely to be systematic, as they relate to technical availability of the marketplaces and are common to us and buyers.


\autoref{table:summarystats} reports the summary statistics for all eight categories. It shows the average unit price (i.e., the price per sales unit in USD), the number of vendors and offers, and the median quantity sold. We use the bitcoin exchange rate on the date of the observation to determine the price in USD.\footnote{While bitcoin is a highly volatile cryptocurrency, in the time frame we study fluctuation in the exchange rate was minimal. We report the exchange rate in \autoref{fig:btcxr}.} Cannabis is the cheapest type of drug on offer (at around 11 dollars per gram), while methamphetamine and heroin are the most expensive (at over 150 USD per gram). Cannabis is also the most popular drug type sold with the most unique offers. The median quantity advertised for the cheapest drugs cannabis (14g) and amphetamine (20g) is significantly larger than for the expensive drugs.

\begin{center}
    [Table~\ref{table:summarystats} about here]
\end{center}

The average price however hides two important sources of price variation: country differences and quantity discounts. The price for the same type of drug, in the same quantity, often shows stark differences by the shipping origin of the product. To illustrate this, \autoref{table:countrydiffcoc} in the appendix documents the price variation for Cocaine across the ten largest countries for the drug, measured by the total number of unique offers. The average price of one gram of cocaine ranges from 81.10 USD in the United States to 283.32 USD in Australia. A likely explanation is that because cocaine must be brought into the country first to be sold from there, differences in the ease of smuggling the merchandise through customs produce very large differences in the cost to obtain the drug.\footnote{Australia operates a  very strict customs regime to protect its unique ecosystem. In addition, cocaine is obtained from the coca plant which requires high moisture and low atmospheric pressure to grow. These conditions are difficult to find or reproduce outside of South America.} 
Most of the offers we observe are from the the western world. \autoref{fig:map} depicts the total number of items observed by the shipping origin country. The largest source of items is the United States at over 10,000 distinct offers, while other countries with a lot of activity in the market are for example the United Kingdom (around 4,600 items), or Germany (around 5,100 items).

The second source of large price variations are quantity discounts. Sellers offer their customers significantly reduced prices for buying in bulk. \autoref{table:discounts} documents the extent of the discounts on offer. Across all categories, sellers continually demand a lower unit price as the quantity bought increases. In the most extreme case, buying 100 grams of amphetamine costs on average only 10\% of the unit price of 1 gram of amphetamine. To account for both country differences and quantity discounts in our estimations, we include seller-, category- and quantity-specific interacted fixed effects which are unique to the country the product is shipped from for all items.\footnote{The handful of country changes in specific items sold we observe are driven by sellers from individual european countries changing the shipping origin to "European Union", as there are no customs procedures or checks between EU member states.}

\autoref{fig:marketsize} shows the number of unique vendor accounts on the two platforms over time. The size of the platforms increased over the latter half of 2014, stabilizing around October for Agora and in December for Evolution. Following the Evolution exit (indicated in grey), the number of vendors on Agora increased as sellers previously present on Evolution sought to continue their business on the only large platform left in the market. However, the size of the increase in vendor accounts indicates that not every Evolution seller re-entered the market following the exit. As we will show in \autoref{sec:Approach}, the vast majority of sellers in the market are single-platform sellers and of those on Evolution, around ten percent move their business onto the other platform. In addition, note that Agora had previously experienced technical difficulties and had more downtime and a reduced speed in accessing the site relative to Evolution. Due to the increased traffic on its site following the exit, the accessibility of the platform suffered further resulting in larger fluctuations of vendors observed in our scrapes. \autoref{fig:uptime} documents the share uptime and speed in accessing the two sites in detail.

\begin{center}
    [Figure~\ref{fig:marketsize} about here]
\end{center}


Although we do have some information on sales, its use is limited. On Evolution, total sales are recorded for each seller. On Agora, sales information is only available in categories that increase in size as the number of sales grows.\footnote{The categories are 0, 1--3, 3--6, 6--10, 10--15, 15--25, 25--40, 40--55, 55--70, 70--100, 100--150, 150--200, 200--300, 300--500, 500--1000, 1000--1500, 1500--2000, 2000--3000, 3000--4000, 4000--5000, 5000+.}
We use a two-step approach to impute sales. We approximate sales using inter/extrapolation of category endpoints for those vendors where we observe changes in categories over time. For those vendors for which we observe only a single category, we rely on linear extrapolation across vendors based on vendors of comparable size. For these reasons, we only use the sales data sparingly. We also have a measure of the number of ratings a vendor has received which we use as an additional proxy for sales throughout the paper, albeit as a lower bound approximation as not every buyer leaves a review after a purchase.

Before proceeding to the analysis, we examine the rating of vendors in more detail. Previous work on the effects of reputation on legal sales platforms has documented ``rating inflation'', i.e.\ that the average rating of a seller tends to be very high in absolute terms and buyers rarely leave a bad review \citep[e.g.][]{tadelis2016, zervas2021, filippas2018}.\footnote{Similar results have also been found for the darknet black market in \citet{bhaskar2017}.} Because conducting transactions in this market requires buyers to reveal their physical address to sellers, this may be further exacerbated due to fears of retaliation. \autoref{fig:rathist} in the appendix shows the distribution of rating across vendors. As expected, the distribution is extremely skewed towards the top on both platforms and exhibits the well-documented ``J-shape'', indicating that the variation in seller rating between (relatively) highly-rated sellers and (relatively) lowly-rated sellers may be quite small in absolute numbers. It appears that when buyers leave a review, most of the time they will tend to leave a perfect or very good review, sometimes a very bad one, but rarely a mediocre one.

\begin{center}
    [Figure~\ref{fig:lifecycle} about here]
\end{center}

\autoref{fig:lifecycle} shows the average rating of a vendor over his/her lifecycle on the Agora platform. We track accounts that have been opened from the day of entry over time. Entry is defined as the date on which the vendor is observed for the first time.\footnote{We also require that the vendor has not made any sales yet, since it is possible for a vendor to be missed in previous scrapes due to technical difficulties. We further exclude the first few scrapes in our dataset when many vendors are observed for the first time.} As vendors mature, the average rating improves and the variation in rating decreases significantly. The improvement in rating becomes significantly less volatile over the first 50-60 days. Within 100 days of activity it appears that sellers on average have matured. However, we continue to observe a slight further increase in the average rating as time progresses. This is particularly true for the small sample of vendors that are active for 300 days and more. The difference in the average rating between a new entrant and a mature vendor is very small in absolute numbers and around 2-3 (percentage) points. \autoref{fig:lifecycle} also indicates that a sizable fraction of new entrants drop out of the market within 80-200 days. The remaining share however continues to trade and its number is stable for a longer time. This suggests that ``good'' sellers stay in the market long-term, while ``bad'' types drop out early on (and may re-enter under a new pseudonym). This is confirmed by the data, as the vast majority of sellers that drop out have a low rating and only few sellers with a low rating continue to stay in the market past 100 days. Since our dataset covers a time period of one year, the number of observations starts to become small and the ratings information more volatile as we track the average entrant for more than 200 days.

We exploit our knowledge of sellers' account names and public PGP keys to link all vendor accounts across both time and platforms. Previous work on darknet marketplaces suggests that only a relatively small fraction of sellers operate across platforms. For example, \citet{soska2015} measure the number of unique aliases (account and marketplace pair) a seller uses and show that more than 75\% of sellers only use one. Similarly, \citet{buskirk2014} suggest that more than 78\% of sellers are only present on a single platform as of September 2014.

\autoref{table:sellers} shows the number of vendor accounts and of operating unique sellers on the two platforms, as well as the number of accounts unique sellers use. There are significantly fewer actual unique sellers in the market than the number of vendor accounts on the two platforms. Of the 2,344 unique sellers active, around 73\% use only one account. This is in line with the previously documented estimates. \autoref{table:sellers} also shows how many accounts a seller that is active on both platforms uses. Almost all sellers use only one account on both platforms respectively, while only 23 sellers use three accounts spread across the two platforms, and three sellers operate with four separate accounts. There are no sellers that only sell on one of the platforms, but use multiple accounts to do so.

For our analysis, we impose further restrictions on the full data to generate a homogeneous estimation sample. As indicated by Figure~\ref{fig:marketsize}, both platforms are growing initially. We restrict the estimation sample to the time period between December 2014 and July 2015, during which platform size is stable. We restrict our analysis to the most frequently traded product: cannabis, cocaine and amphetamine. We exclude MDMA from the analysis. MDMA is sold both in powder and in pill form (``Ecstasy'') with varying advertised strength. Due to the difficulty of separating pill listings from powder and the fact that pill dosage is generally exaggerated by sellers, it is impossible to generate a uniform measure of quantity without incurring substantial measurement error. For the remaining categories, we do not have sufficient coverage of sellers on both platforms within country-quantity cells.
Similarly, we exclude offers for which the country is unknown and where coverage across both platforms within the country across all three products is insufficient. Most of these are developing countries which have only a limited number of sellers and domestic demand\footnote{Many items in these countries appear to be listed primarily for export only (sometimes explicitly excluding domestic buyers). Others have strict customs regimes limiting cross-country competition. We exclude the following countries: Undeclared, Australia, South Africa, China, New Zealand, India, Philippines, Pakistan, Thailand, Colombia, Guatemala, Andorra, Cambodia, Peru.}. To smooth some of the noise out of the daily data, we aggregate the data in weekly intervals.




 \section{Empirical approach}
 \label{sec:Approach}
 To evaluate the effect of the platform closure on prices, we conduct an event study analysis. We define an individual item as the unique offer observed on one of the platforms, sold by one specific seller, belonging to one drug category, and of a given quantity. We denote the individual items by the index $i$ and time by $t$. We estimate a conventional event study specification,
\begin{gather}\label{event-eqn}
	\log(\text{Price}_{it}) = \alpha_{i} + \delta_{t} + \sum_{\tau}\beta_{\tau} \text{d}^{\tau}_{it} + \gamma\ \text{Escrow}_{i} + \varepsilon_{it} \ , 
\end{gather}
where $\text{Price}_{t,i}$ denotes an item $i$'s unit price at time $t$, $d_{it}$ measures exposure to the exit-scam relative to event time $E_{i}$ for Evolution sellers, $d_{it}=\mathbbm{1}\{t - E_{i} = \tau \}$, and $\beta_{\tau}$ measures the effect of the exit on an item's price $\tau$ periods relative to the platform exit. $\text{Escrow}_i$ measures whether the item requires use of the escrow system.\footnote{In our estimations, we consider specifications with and without the escrow covariate. Inclusion of the variable does not affect the results.} $\delta_{t}$  and $\alpha_{i}$ are time- and unit-specific fixed effects. $\varepsilon_{it}$ is a scalar unobserved item-specific shock at time $t$ that is assumed to be mean-independent of the remaining right-hand side variables.

We exclude the last week pre-exit as the reference period, normalizing $\beta_{-1}=0$ and cluster standard-errors at the vendor-substance level. As outlined in the previous section, we restrict our data to a homogeneous period during which platform size is stable. We consider an event window of 14 weeks pre-exit and 16 weeks post-exit. Due to a six-week period of missing data during which scrape information is unavailable, we aggregate all pre-treatment periods prior to the data gap in the first pre-treatment coefficient after the gap for visualization purposes. This implicitly assumes constant treatment effects prior to $\tau=-5$ (but is inconsequential for the results).

We specify the unit fixed effects at the lowest possible level, i.e., the item (vendor-substance-quantity) level. In contrast to separating the individual components, this specification exploits the shift in the most flexible way, controlling for both seller- and item-specific time-constant effects and unobserved product quality. By conditioning on quantity in the item-specific fixed-effects, we also explicitly allow for non-linear pricing of products. We documented previously in \autoref{sec:Descriptives} that quantity discounts are commonplace. Note that we implicitly account for country of origin without explicitly conditioning on it, as we do not observe sellers offering the same product in multiple locations. Finally, the weekly time fixed effects control for overall shifts in demand or to exchange rates that are common to all offers.


In addition to the reduced-form event study approach, we also present estimates of the effect of rating on price for Evolution sellers forced to re-enter the market. We consider the following pricing equation:
\begin{gather}\label{main-eqn}
	\log(\text{Price})_{it} =  \alpha_{i} + \delta_{t} + \beta \log(\text{Rating})_{jt} + \gamma\ \text{Escrow}_{i} + \varepsilon_{it}, 
\end{gather}
where $\text{Rating}_{jt}$ is seller $j$ of item $i$'s aggregate rating at time $t$. Since ratings can be zero, we use the inverse hyperbolic sine transformation instead of the natural logarithm to normalize the rating variable \citep{burbidge1988,bellemare2020}.

Since rating information is a summary measure of past buyers feedback, it is likely to be a function of past prices. Buyers who purchase an expensive product may have a correspondingly higher expectation of its quality, which will impact the rating they leave. Then the ratings variable is correlated with past realizations of the error term. To overcome these endogeneity concerns, we exploit the fact that the Evolution platform performed an exit-scam during the time frame we study. By linking seller accounts over time and across platforms, we are able to track sellers that are forced to re-enter the market by registering on the Agora platform due to the exit-scam of the Evolution platform.
We then exploit the forced re-entry and ratings reset of Evolution sellers to estimate the effect of rating on price. We augment our pricing equation with the following first-stage regression for an item's rating,
\begin{gather}\label{first-stage-eqn}
	\log(\text{Rating})_{jt} = \mu_{i} + \eta_{t} + \phi \mathbbm{1}\{t \geq E_{i}\} + \lambda \text{Escrow}_i + \xi_{it}. 
\end{gather}

We track the evolution of ratings over time to verify both that platform exit depresses the rating of re-entering sellers and that the average ratings of sellers on Agora and migrating sellers evolve similarly. The result is shown in Figure~\ref{fig-rating-time}. We find that ratings across both groups are stable before the exit scam. After the exit, there is a sharp drop in the rating and a subsequent recovery. The average rating of sellers that re-enter the market recovers within the three months following the exit, which is in line with the approximately 100 days it takes for the average seller to mature. The rating of sellers using Agora shows no reaction to the exit. Re-entering sellers however have a higher rating than the average seller prior to exit. We consistently observe in our data that re-entering sellers are drawn from the upper part of the ratings distribution pre-exit.


We document the effect on rating in more detail in \autoref{fig:sw-histdrop}, that depicts the placement of re-entering Eolution sellers in the ratings distribution prior and post exit. Prior to the exit, the highest share of migrating sellers was in the top bar of the distribution with a rating of 99.5 or above. This corresponds to being in the top 50\% of the ratings distribution, however most of these sellers have ratings that put them in the top 25\% of the distribution. In the month following the exit on the other hand, the highest share of Evolution sellers are observed in the distribution at 93-93.5, corresponding to being in the bottom 15\% of the distribution and a much smaller share is found in the upper parts of the distribution.


\begin{center}
	[Figure~\ref{fig:rating-time} about here]
\end{center}



We also observe a corresponding drop in estimated sales for Evolution sellers after the exit (Figure~\ref{fig:shock-sal}). The sales growth of Agora sellers is again unaffected by the exit. Both before and after the exit, sales of both groups grow at a very similar rate. Since some of the sales information we have is imputed, we also proxy sales by the number of feedbacks left for a seller and observe the same finding (see \autoref{fig:shock-fb} in the appendix). Both sales and feedback received evolve in parallel between the two groups of sellers before and after the shock.

\begin{center}
	[Figure~\ref{fig:shock-sal} about here]
\end{center}


Identification relies on an exclusion restriction for Evolution sellers who re-enter. We assume that migrating only affects prices via the reset in rating, ruling out direct effects of the platform move on price. However, this assumption need only hold conditionally within item types, i.e., within homogeneous classes of drugs, for a specific seller, and a given quantity. Confounding factors need to be both time-variant and not absorbed by the fixed effects. Importantly, this implies that we can disregard seller-specific changes in pricing strategies that affect all items of a seller equally, or indeed changes in pricing strategies that are unique to the category of drug for a specific seller.

One possible change in pricing strategy that could potentially confound our interpretation of the rating effect is if sellers react to changing platforms by adjusting prices to recapture market share. In this case, the effect of rating on price would be driven by strategic pricing behavior, but as just discussed this would only confound our interpretation if the strategic behavior is item-class specific. While we cannot test for this possibility directly, we do not observe any change in the rate of sales made by Evolution sellers from before to after the exit (see \autoref{fig:shock-sal}), nor in the rate of feedback left (see \autoref{fig:shock-fb}) that would indicate a fast recapturing of market share of Evolution sellers. In addition, recent work documents that firms do not make use of markups to increase their market share \citep{fitzgerald2018}. This is further reinforced by the institutional structure of the market and the high degree of competition we observe.

\begin{center}
	[Table~\ref{table:sw-prevspost} about here]
\end{center}

\autoref{table:sw-prevspost} documents the offers and prices of re-entering Evolution sellers fourteen days before and fourteen days after the exit. The figures suggest that former Evolution sellers do not vary their product portfolio. Both the average number of offers and the median quantity offered remain stable. However, the figures suggest that sellers adjust their prices. The average unit price of Evolution sellers between the two dates decreases across all categories, suggesting an adjustment to the reputational loss suffered. The percentage change of prices is significant for all drugs and is approximately -5.3 USD on average.  The market price change on the other hand between fourteen days prior to the exit and fourteen days after is stable or slightly increasing for most categories, reinforcing that losing reputation has a negative effect on the prices a seller may charge. We also test whether the unit price change for Evolution sellers vs.\ the remainder of the market is significant. The p-values suggest that the difference is significant for all products and Cannabis, the largest category, while bordering conventional significance levels for the two remaining categories.

Lastly, \autoref{table:sw-pre-exit} in the appendix contrasts re-entering Evolution sellers to all other sellers present on Evolution a week prior to the exit. It demonstrates that these Evolution sellers are representative for the average Evolution seller prior to exit and found in almost identical proportion across the different categories compared to the average seller. Re-entering sellers tend to offer slightly fewer different products on average than other Evolution sellers across most categories, while the median quantity of items offered is similar between the two groups. These patterns fit with the higher average rating we observe: Former Evolution sellers tend to be reputable, highly rated sellers that offer a somewhat smaller, higher quality portfolio compared to the average seller, but are present in the different product groups in the same share as the average seller and offer the same quantities as the average seller. That is, they are a representative sample of the high-quality sellers of the entire market. Indeed, when comparing re-entering sellers only to other highly-rated sellers, the difference in the number of offers decreases and the pattern disappears. This aspect provides a unique interpretation of our results that sheds light on the value of information reveal about seller types in a market with quality uncertainty, since former Evolution sellers can on average be considered revealed high types prior to the exit and entrants whose type is uncertain following the exit.


\section{Results}
\label{sec:Results}
\autoref{fig:eventstudy} shows our estimates of the effect of re-entering the market following the Evolution exit on unit prices charged based on the event study specification~(\ref{event-eqn}). We find the same pattern for all categories of drugs. The forced re-entry has a negative, statistically significant effect on prices over the first 12-13 weeks following the Evolution exit. The estimated effect gradually increases in absolute value and peaks at 8-9 weeks after the exit consistently across the three categories of drugs. For the full sample we find that unit prices were almost 10\% lower after 7-9 weeks. Our separate estimates for cannabis and amphetamine are slightly smaller, while those for cocaine are largest at 13\%. These differences are explained by the difference in price levels among types of drugs---cocaine is significantly more expensive (see \autoref{table:summarystats}). The effect disappears after 12-13 weeks as sellers that re-entered the market recover their rating. We documented previously in \autoref{sec:Descriptives} that it takes around 100 days for a new entrant to establish themselves as a highly-rated seller.

\begin{center}
	[Figure~\ref{fig:eventstudy} about here]
\end{center}


These price decreases represent a significant revenue loss. For example, over the course of weeks 7-10 following the exit, these sellers on average obtain approximately 34 new individual reviews. Prior to the platform exit, the average unit price for cannabis sold by sellers that changed platforms was 12 USD, while the median quantity was 10 grams. Given the effect we find and using number of reviews to approximate sales, this suggests that sellers of cannabis lost on average approximately 230 USD over the course of four weeks. As not every buyers will leave a review when purchasing, the number of individual reviews is likely a downward biased estimate of sales, suggesting a larger loss.

We have a coarse measure of the number of deals made by a seller in our data that we use to impute the approximate sales made (see \autoref{sec:Descriptives} for an explanation of our sales information). During weeks 7-10 following the Evolution exit we estimate that 59 sales were made on average resulting in a revenue loss for sellers of cannabis of approximately 400 USD during weeks 7-10 after the Evolution exit. For sellers of cocaine the income loss at prices 12\% lower than prior to the exit is approximately 1,360 USD, while for sellers of amphetamine the loss is 430 USD\@. Note that as we cannot relate the individual reviews or sales to individual items, we assume for all our revenue calculations that all reviews/sales for a given seller relate to the same category of drug.

We can also obtain an estimate of the total revenue loss experienced by sellers. They obtained on average 210 new reviews in the 12 weeks following the Evolution exit. Given the average effect size over the same time frame, we calculate that for the full sample, based on the number of new reviews, the total revenue loss for sellers is 2,900 USD\@. For sellers of cannabis, we calculate the loss to be approximately 900 USD, for sellers of cocaine 4,100 USD, and for sellers of amphetamine 720 USD\@. The same calculation based on the number of new (imputed) sales shows a revenue loss of 4,800 USD for the full sample, 1,500 USD for cannabis, 7,000 USD for cocaine, and 1,170 USD for Amphetamines.

\begin{center}
	[Table~\ref{tab:fs} about here]
\end{center}

Next, we proceed to isolate the effect of rating on price. \autoref{tab:fs} presents the estimation results for the first stage regression outlined in equation~(\ref{first-stage-eqn}) in \autoref{sec:Approach}. We find a consistent negative effect of the market re-entry on log rating with an estimate of approximately $-0.09$. Thus, we find that re-entering leads to a roughly 9\% decrease in rating. The overall estimate is statistically significant at the 5\% level, while the estimates for the individual categories are insignificant. We attribute this to a lack of power, as the number of re-entering sellers we observe is smaller for individual drug categories. This is reinforced by the fact that the estimates remain stable while the standard errors increase consistently with decreasing sample size. We report $p$-values for transparency.

We further find a positive estimate for escrow usage on rating for all drugs of $0.018$ that is statistically significant at the 10\% level. The effect size is again similar across the individual categories of drugs but statistically insignificant. We would expect a positive effect: Since the escrow system is meant to reduce the risk buyers face when paying for merchandise before receiving it, use of the system is appreciated by buyers and associated with a higher rating.

\autoref{tab:iv} presents the estimation results for the pricing model outlined in equation~(\ref{main-eqn}), where we exploit the exogenous variation in rating induced by the forced re-entry of Evolution sellers to estimate the ratings effect on price. We find a strong, positive and statistically significant effect of rating on price for our full sample. Our estimates for the individual categories of drugs are also substantial but again statistically insignificant. The estimates consistently become more noisy across drug categories with smaller sample size. Due to the log-log specification, our estimate can be interpreted as an elasticity: increasing rating by one percent causes an increase in the unit price of 1\%. At the average pre-treatment unit price of 36.5 USD, this corresponds to a change in price of 0.37 USD\@. The standard deviation of log(rating) prior to the exit is approximately 0.12, implying sellers that improve their rating by one standard deviation may charge 12\% higher unit prices or 4.40 USD more on average. This is equivalent to an increase of almost 10\% of one standard deviation of unit prices.


\begin{center}
	[Table~\ref{tab:iv} about here]
\end{center}



These findings represent an economically significant value to being identified as a reputable seller. As before, we can make use of our information on obtained reviews and imputed sales to estimate the revenue loss due to the loss in rating that Evolution sellers suffered. As we cannot ascertain the proportion of sales associated with individual items offered by a seller, we assume that it is uniformly distributed. At the end of the 10 week post-exit estimation window, sellers that re-entered the market had on average obtained about 185 new reviews and a rating about 2.5 points lower. Prior to the exit, the average quantity offered by these sellers was 7 grams and the average unit price was 36.5 USD\@. Based on our estimate, we therefore calculate that on average they suffered a revenue loss of over 1,000 USD in the 10 week timeframe post-exit as a consequence of a lower rating. The income loss becomes even more stark when considering our measure of sales. We estimate that former Evolution sellers made 307 new sales on average in the 10 week post-exit estimation window. Based on this, these sellers on average had a reduction in revenue of close to 1,800 USD solely due to the lower rating and reputation.

We do not find any effects for offering escrow services on prices in Table~\ref{tab:iv}. This is likely due to the small number of items that change their escrow state. Since we include item fixed effects in our estimation, the escrow coefficient is identified from items that switch their offering from non-escrow to escrow, which happens very rarely. Most sellers offer the same item in escrow and non-escrow versions in parallel, and descriptive statistics suggest a price differential between these offers.

As the sellers that we track on average were highly-rated, reputable sellers prior to the Evolution exit but new entrants with no reputation after the exit, these calculations can be interpreted as the monetary value of being an identified ``good'' type in a separating equilibrium, instead of a pooling one in a setting with unobservable quality.  During an average week prior to the exit, the sellers we track obtained 40 new reviews, and their average rating prior to exit was 8.8\% higher than immediately after re-entry. Thus, based on the number of reviews, we calculate that on average the value of being in a separating equilibrium is close to 900 USD per week for good types. The same calculation using our sales information in turn implies additional revenue of 1,650 USD per week.

However, as sellers recover their rating following the exit period, the impact of the ratings shock should decrease. Our event study analysis shows that the effect on prices begins to decrease after roughly 9-10 weeks and our descriptive analysis suggests that it takes about 100 days for a seller to establish themselves as a reputable supplier. We re-estimate our instrumental variable specification by extending the post-exit estimation window one week at a time. The estimated coefficient and confidence interval is shown in \autoref{fig:effect-over-time}. Our findings are exactly in line with what we expect. The estimated effect continuously increases over the first 10 weeks post-exit and then gradually decreases as the former Evolution sellers' reputation recovers.

\begin{center}
	[Table~\ref{fig:effect-over-time} about here]
\end{center}


We also present a series of sensitivity and robustness checks. As a first sensitivity analysis, we repeat our main specifications when excluding the escrow covariate. The results are shown in tables~\ref{tab:fs-nc} and~\ref{tab:iv-nc} in the appendix. The estimates are unaffected by exclusion of the variable.

In \autoref{tab:ols}, we also present estimation results for a naive estimation approach of the effect of rating on price using OLS\@. We find a small, statistically significant effect of rating on price for cannabis, but not for cocaine, amphetamine, or the full sample. This effect implies that a change in rating by one percent is associated with a change in the unit price by 0.13\%. These findings are broadly in line with previous literature on legal marketplaces that suggests that there are only small or no effects of rating on price \citep[e.g.][]{resnick2002,cabral2010,houser-wooders-06,cai2014}. Our overall estimate of 0.006 in \autoref{tab:ols} is also in line with previous work on darknet marketplaces by \cite{espinosa-2018}, who investigates a different platform and notes that a 10\% increase in positive evaluations is associated with 0.3-0.6\% higher prices.
Our results incorporating the exogenous variation in rating induced by the platform exit in \autoref{tab:iv} indicate that there is a larger price premium for having a higher reputation than previous estimates suggest.

Finally, \autoref{tab:pl} presents estimation results for a placebo exit in the period prior to the actual platform exit. We assume a placebo exit takes place 11 weeks prior to the actual Evolution exit. We then consider an estimation window of 15 weeks prior to the placebo exit and 10 weeks after. This is equivalent to the estimation window we consider around the real exit. We find no statistically significant effects of rating on price, instead all of the estimates are very noisy. 



\section{Conclusion}
\label{sec:Conclusion}
In this paper, we analyze the value of reputation in a market devoid of legal institutions. We examine the online market for illegal drugs, where market participants transact on online platforms that provide a rating system for sellers in a form familiar to any user of popular legal platforms such as Amazon, thus providing a publicly observable measure of reputation. The lack of legal institutions and strong need to stay anonymous suggests that reputation is the sole force to facilitate trade among market participants. We make use of a novel dataset of webscrape information of offers on the two dominant platforms during 2014/2015 that jointly covered more than 90\% of the market. We exploit the fact that one of the two platforms suddenly exited the market in March 2015 and track sellers that were forced to migrate to the remaining platform. By necessity, these sellers must open a new account to continue selling, thereby resetting their rating. We evaluate the effect of market re-entry and exploit the induced exogenous variation in rating to identify the effect of reputation in the form of online ratings on the price a seller may charge in the absence of legal institutions.

We document three main results. One, our event study indicates that re-entering the market leads to a decrease in prices of up to 9\% on average. The effect on prices grows gradually and peaks around 7-9 weeks after exit. After 12-13 weeks the effect has disappeared, which is in line with the time it takes for an entrant to establish themselves as a highly-rated seller. Two, we exploit the exogenous variation in rating due to re-entering the market to estimate the effect of rating on price and find an elasticity of about one. We show that once again this effect begins to decrease after approximately 10 weeks following the exit. Three, we document that re-entering sellers were representative of highly-reputable sellers prior to the platform exit and of other entrants following the exit. We thus calculate that the weekly additional revenue for a good type playing a separating equilibrium instead of a pooling one in a setting with unobservable quality and no legal institutions is approximately 1,650 USD\@. Finally, our empirical approach implicitly shows that a sellers reputation is at least partially non-transferable across online marketplaces.


Our work in this paper corroborates previous literature which suggests that reputation plays an important role in facilitating trade when governmental or legal institutions are lacking. To the best of our knowledge we are the first to explicitly estimate the value of reputation in such a unique setting. We also add to the large literature on reputation on online sales platforms by investigating reputational effects for illegal products. Given the abundance of black markets for many types of goods, studying the dynamics of reputation in more detail in such an institutional void is a promising pursuit for future research.


\clearpage
\appendix
\setcounter{table}{0}
\setcounter{figure}{0}

\section*{Main Figures and Tables}
\label{sec:mainresults}

\begin{figure}[!h]
  \centering
  \caption{Platform size}
  \includegraphics[width=\linewidth]{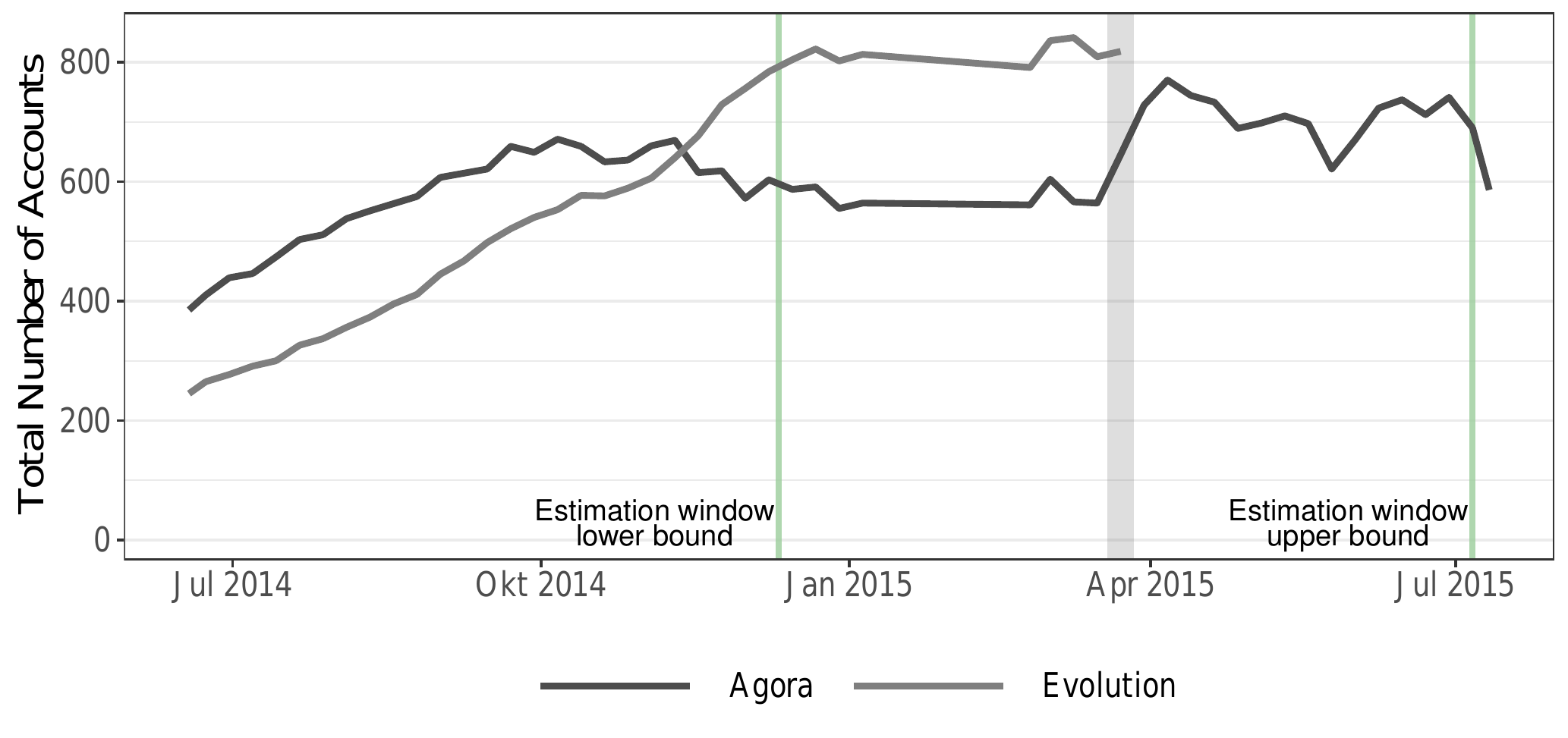}
  \label{fig:marketsize}
  \vspace{-0.5em}
  \fnote{\textbf{Notes:} The figure shows the number of unique vendor accounts active on the two platforms. The Evolution exit is indicated in grey. The flat lines in early 2015 are due to missing data. Accessability of the Agora platform deteriorated in particular after the exit (see \autoref{fig:uptime} in the appendix).}
\end{figure}

\begin{figure}
  \centering
  \caption{Vendor lifecycle}
  \includegraphics[width=\linewidth]{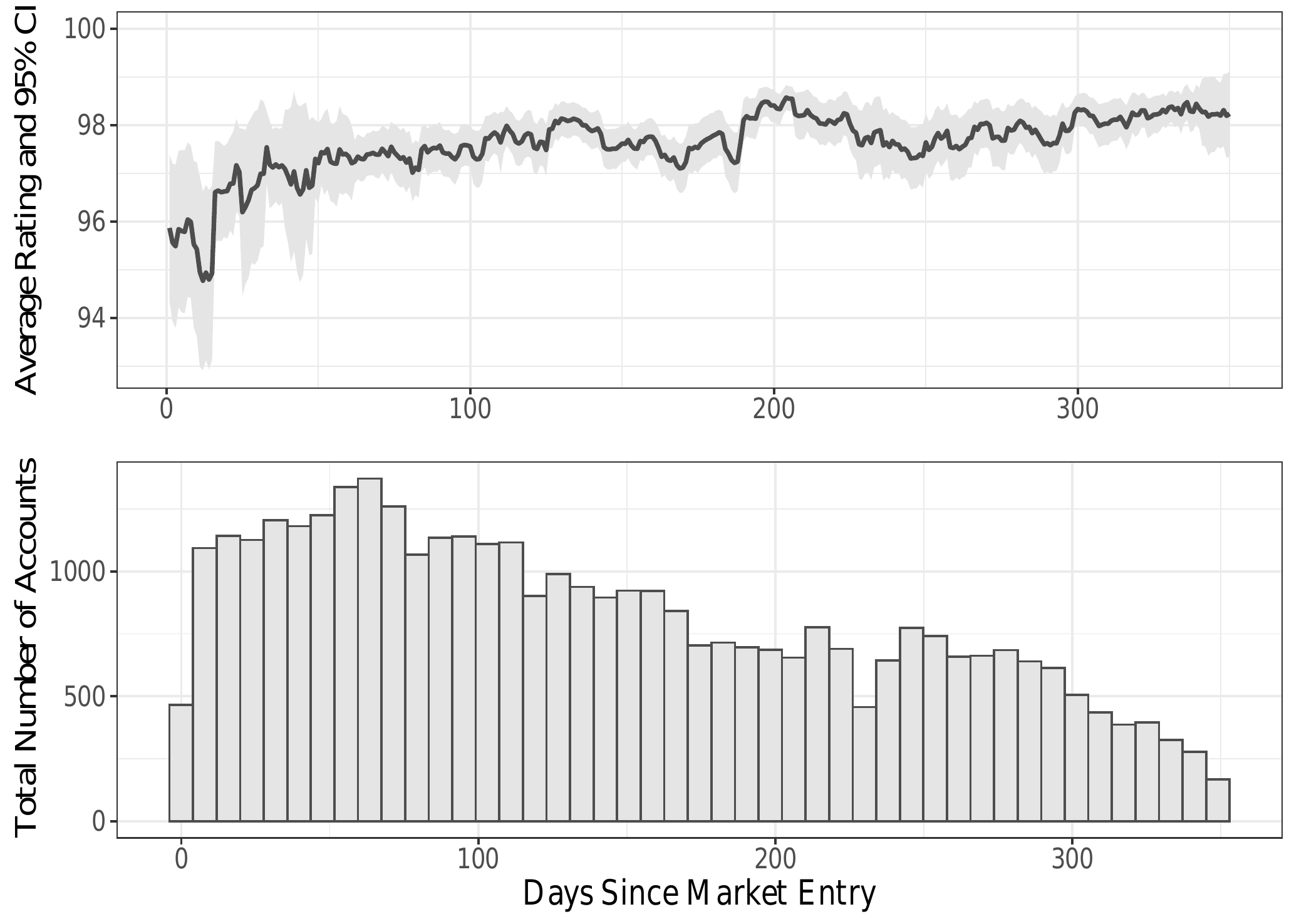}
  \vspace{-0.5em}
  \label{fig:lifecycle}
  \fnote{\textbf{Notes:} The figure shows the mean rating at the top  and the number of unique vendor accounts at the bottom observed by the number of days passed since the vendor entered the market. Rating is measured on a scale of 0 to 100 with higher numbers indicating better rating. Entry is defined as the first date of observation for the account. We exlude accounts of sellers that have already made sales before the first time they are observed. The 95\% confidence band of the average rating is shown in grey.}
\end{figure}

\begin{figure}
  \caption{Evolution of ratings over time}
  \label{fig:rating-time}
  \includegraphics[width = \linewidth]{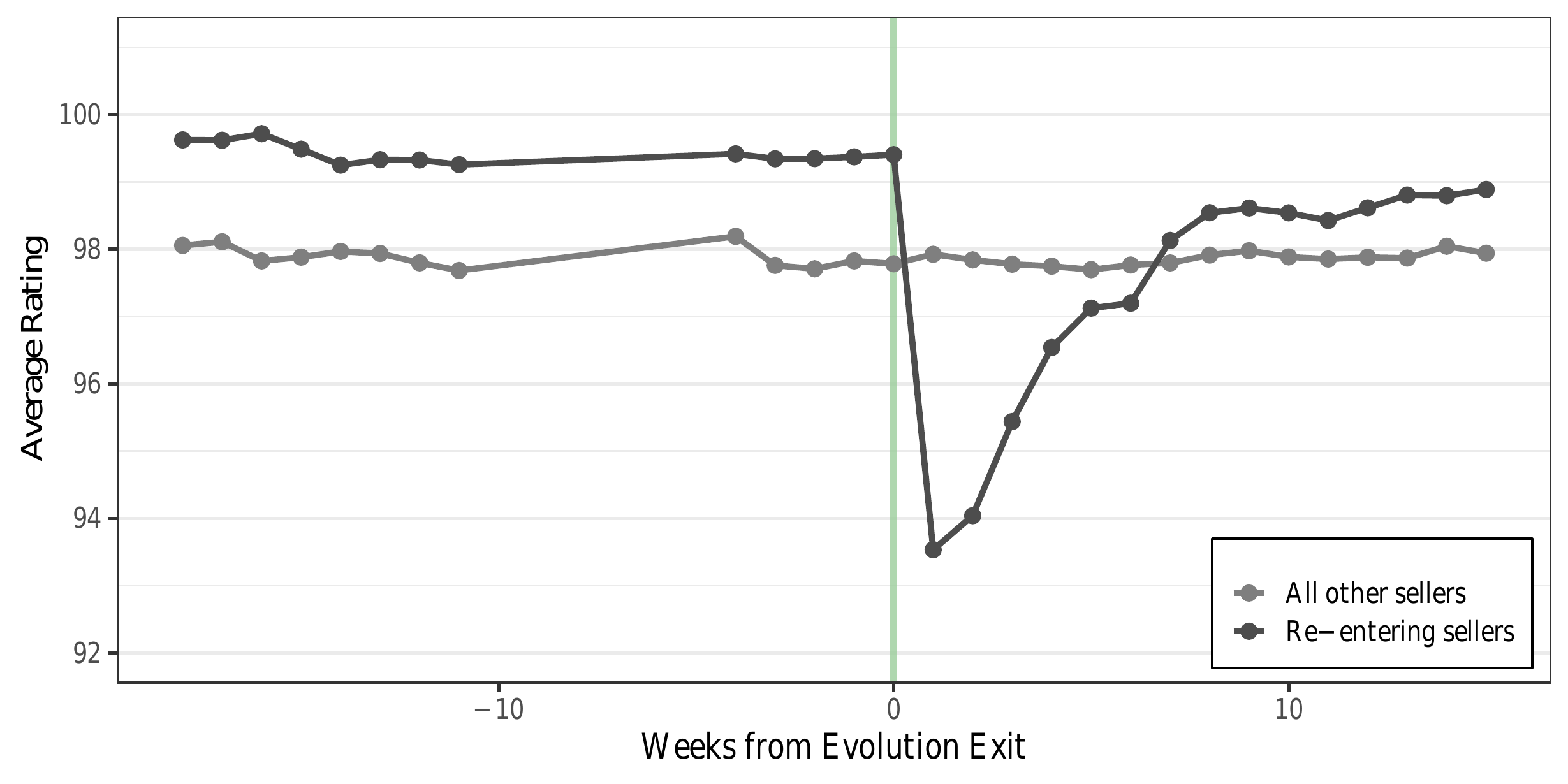}
  \fnote{\textbf{Notes:} The figure shows the mean rating of re-entering Evolution sellers and all other sellers. Rating is measured on a scale from 0 to 100.}
\end{figure}

\begin{figure}
  \centering
  \caption{Evolution of sales over time}
  \label{fig:shock-sal}
  \includegraphics[width=\linewidth]{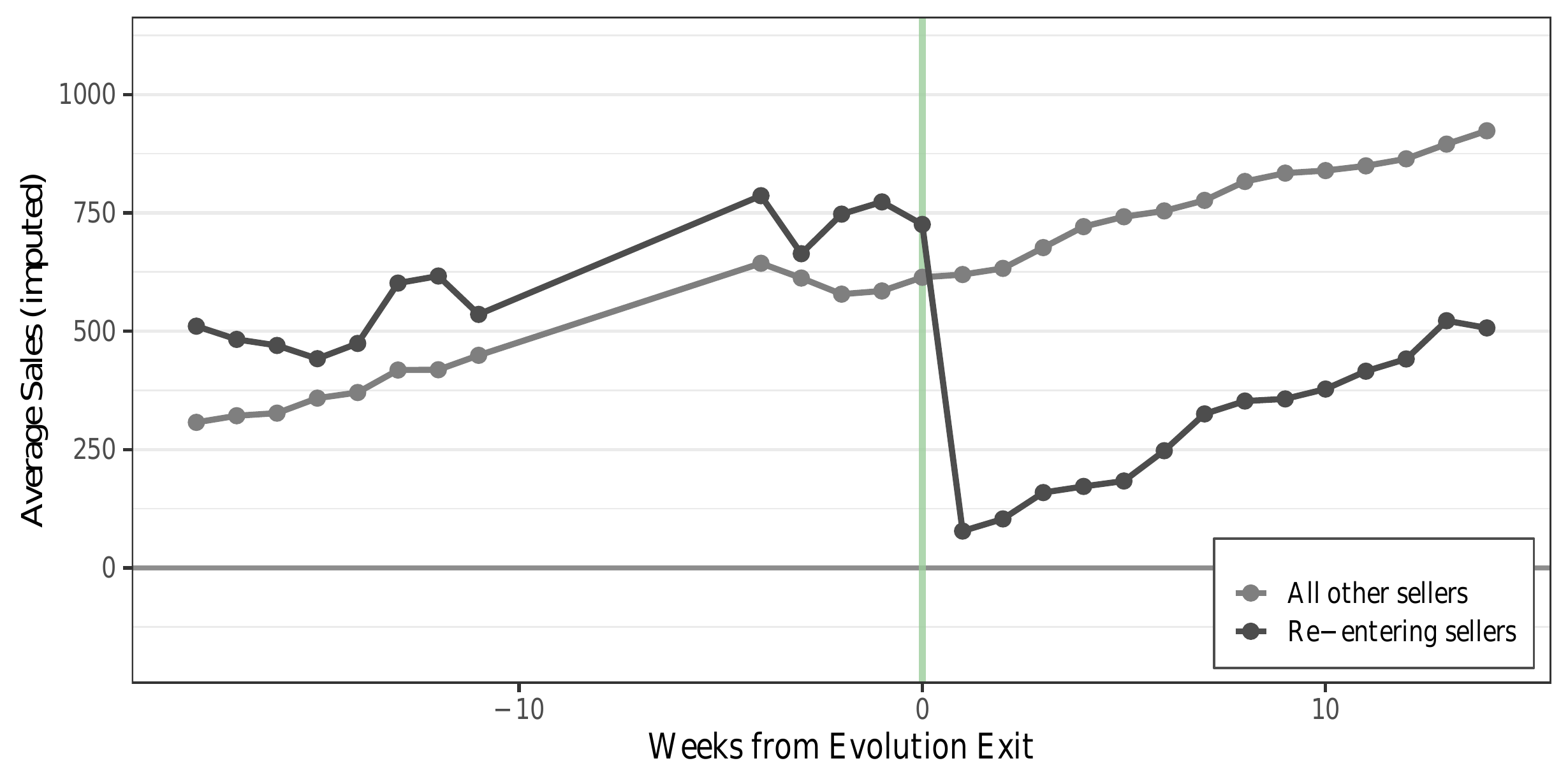}
  \fnote{\textbf{Notes:} The figure shows the average sales of re-entering Evolution sellers and all other sellers. The Evolution exit period is highlighted in green. Sales are imputed for Agora as outlined in the data section.}
\end{figure}

\begin{figure}
  \centering
  \caption{The effect of platform exit and re-entry on prices}
  \label{fig:eventstudy}
  \begin{subfigure}{0.49\textwidth}
    \caption{All}
    \includegraphics[width=\linewidth]{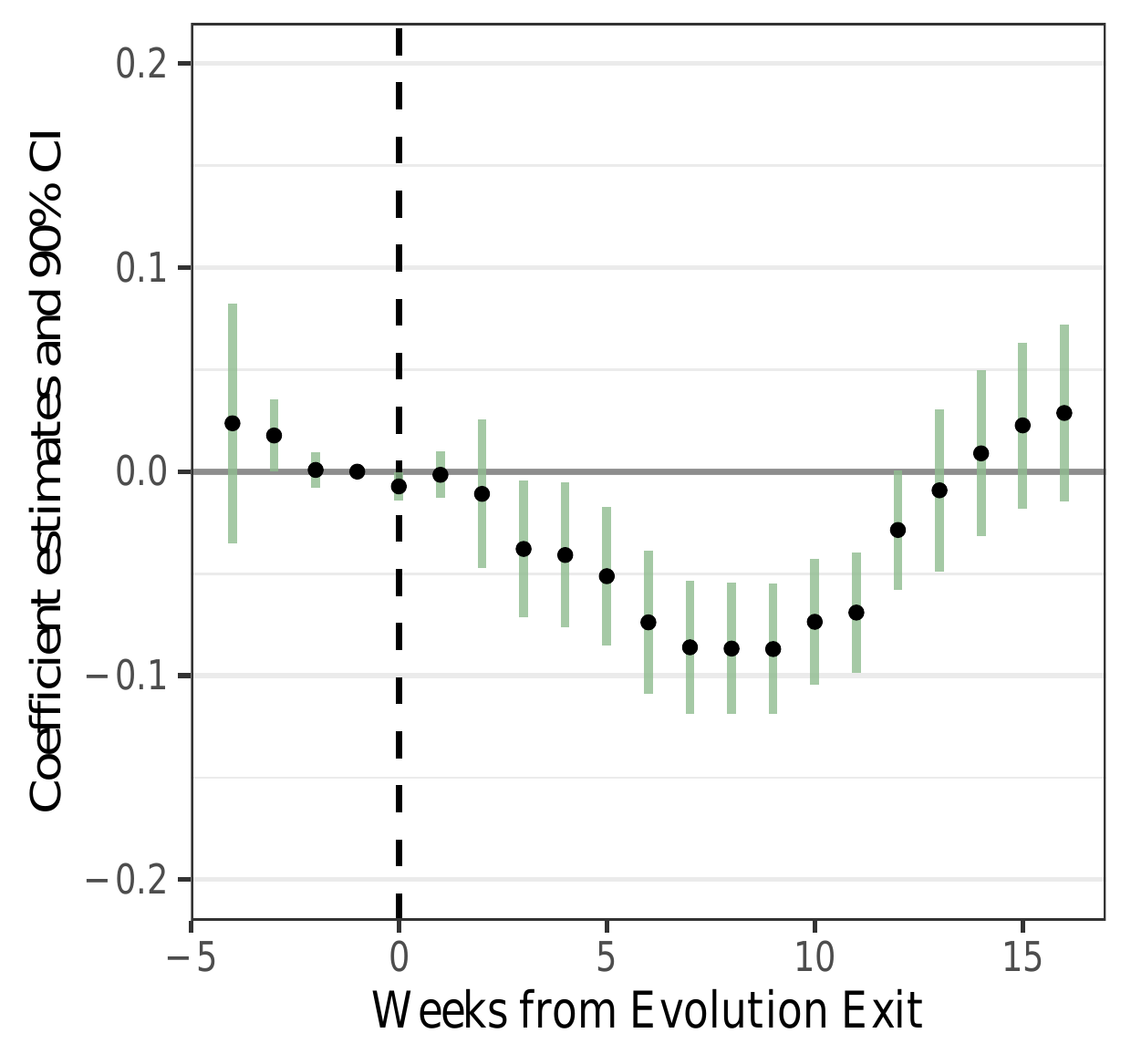}
  \end{subfigure}
  \vspace{2em}
  \begin{subfigure}{0.49\textwidth}
    \caption{Cannabis}
    \includegraphics[width=\linewidth]{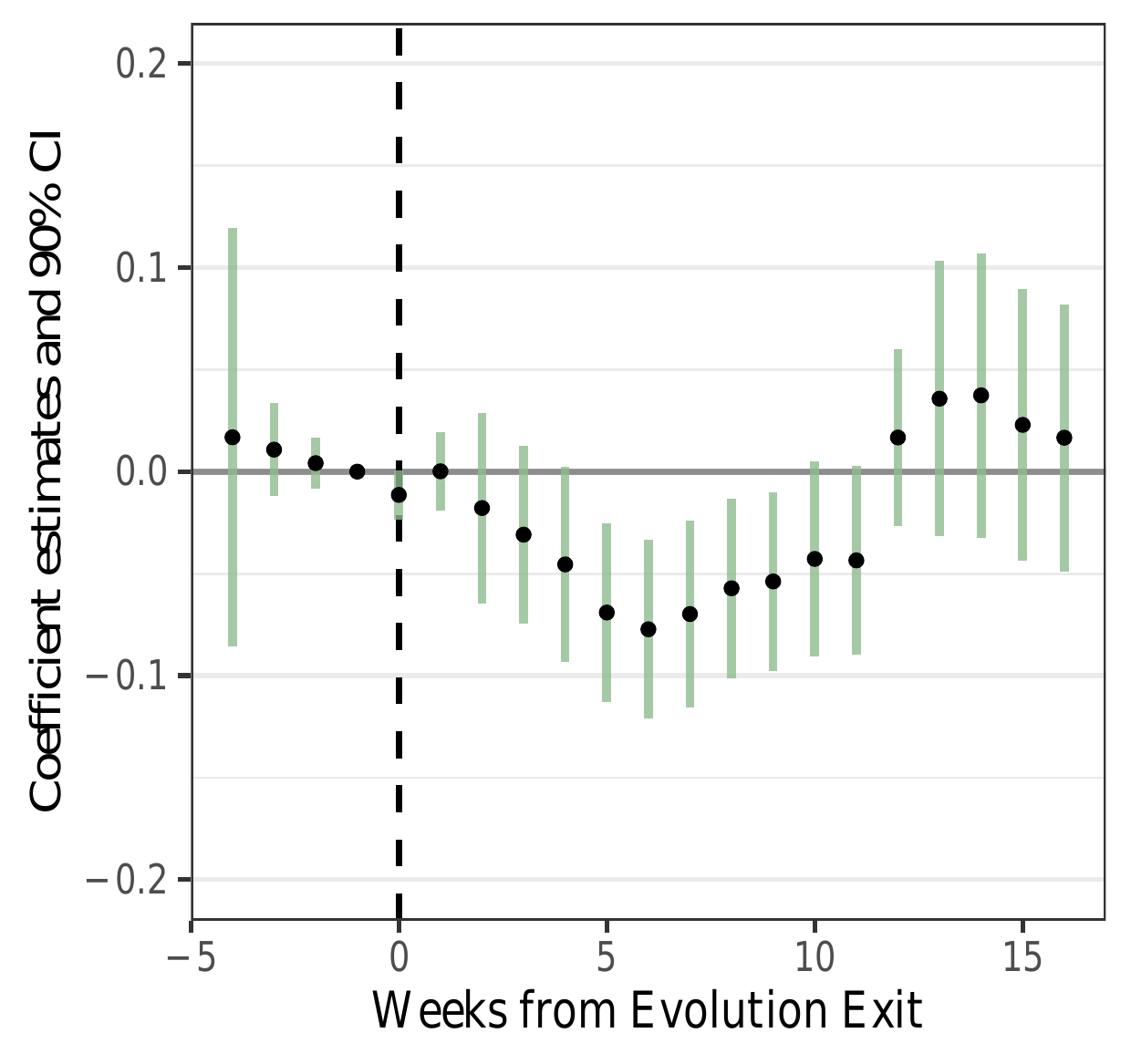}
  \end{subfigure}
  \vspace{2em}
  \begin{subfigure}{0.49\textwidth}
    \caption{Cocaine}
    \includegraphics[width=\linewidth]{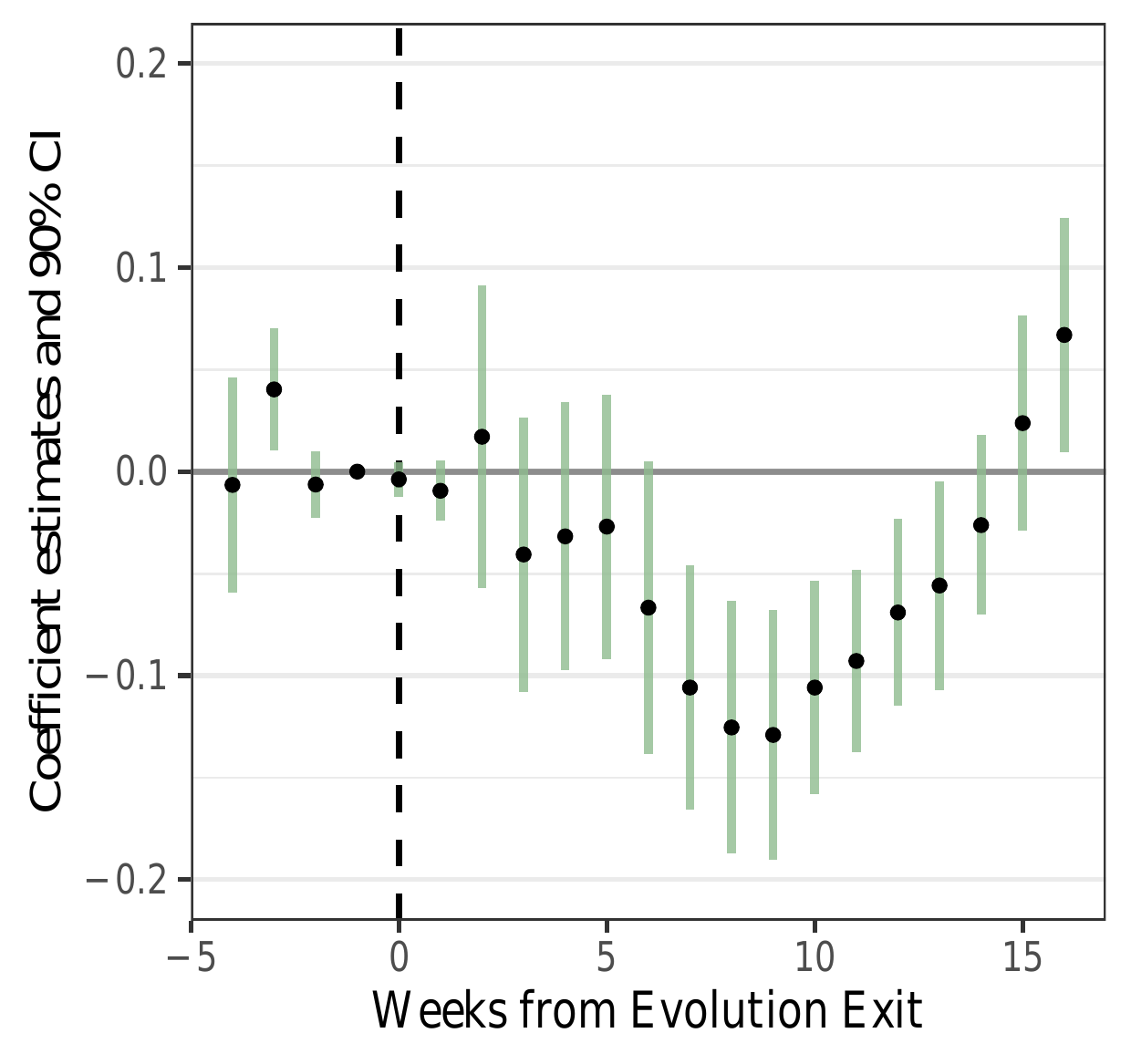}
  \end{subfigure}
  \begin{subfigure}{0.49\textwidth}
    \caption{Amphetamine}
    \includegraphics[width=\linewidth]{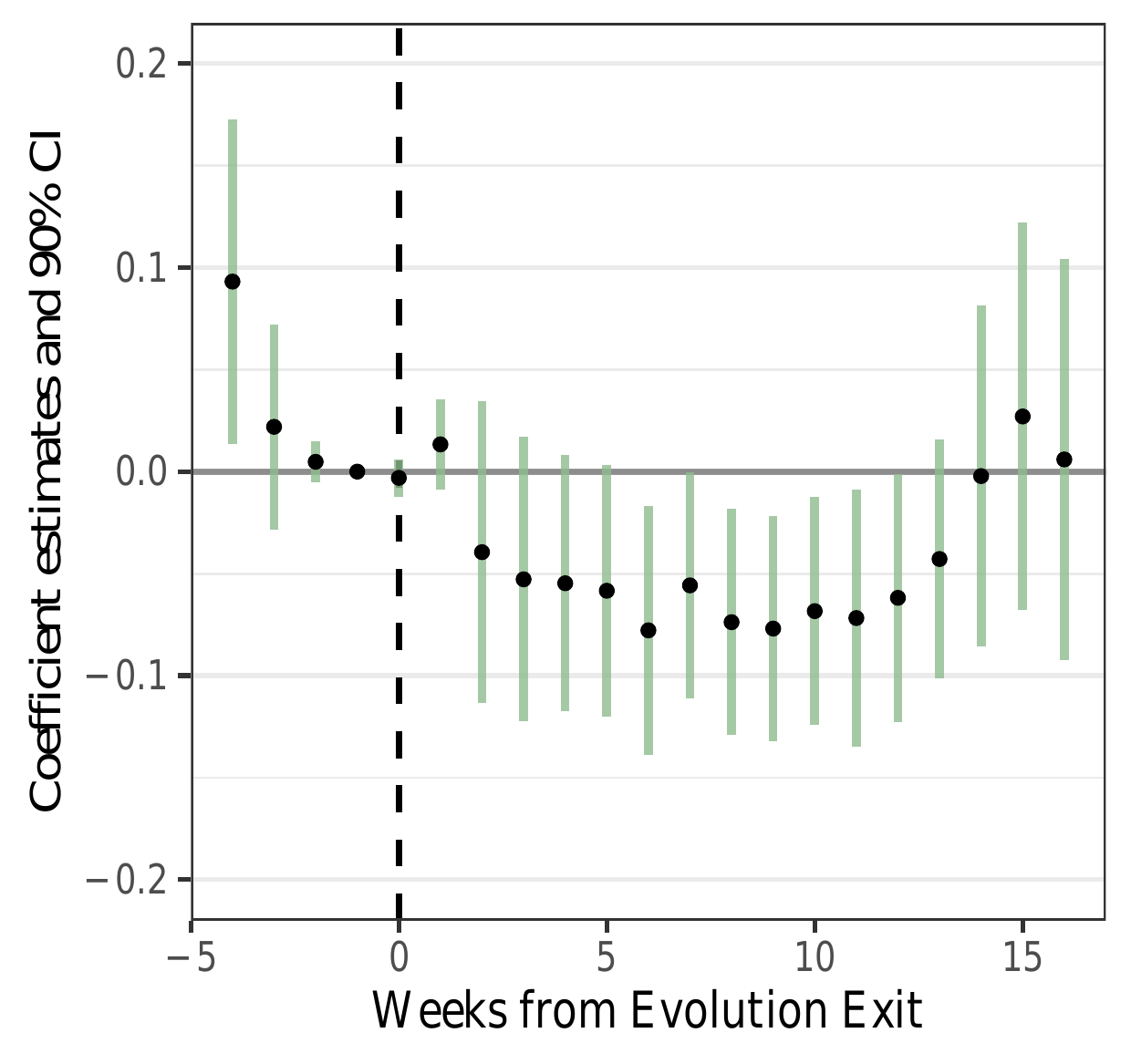}
  \end{subfigure}
  \fnote{\textbf{Notes:} The figure shows the estimated effect of week dummys on the logarithm of unit price charged. Dots correspond to the coefficient, with confidence intervals shown in green. Standard errors are clustered at seller-substance level. Item-level fixed effects are included. The full details of the specification are found in \autoref{sec:Approach}.}
\end{figure}

\begin{figure}
  \centering
  \caption{The effect of rating on price over time}
  \includegraphics[width=0.9\linewidth]{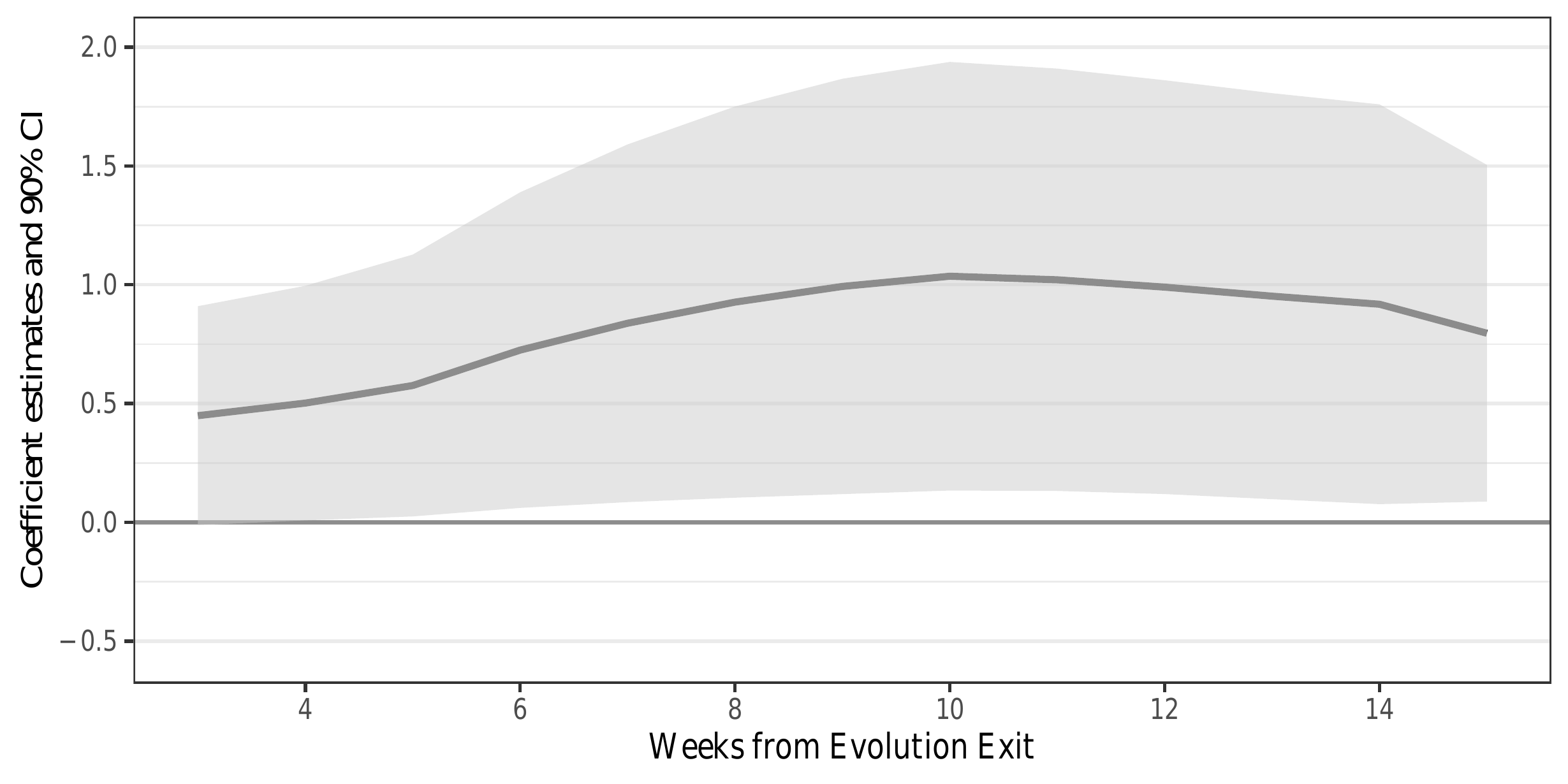}
  \label{fig:effect-over-time}
  \fnote{\textbf{Notes:} The figure shows the estimated coefficient of rating on unit price based on a linear model as specified in \autoref{sec:Approach} with weekly data. The time frame post-exit is extended one week at a time and the effect re-estimated. The 90\% confidence interval is shown in grey.}
\end{figure}

\clearpage


\begin{table}[ht]
	\centering
	\captionsetup{width=\textwidth,justification=raggedright,singlelinecheck=off}
	\caption{Summary statistics}
	\label{table:summarystats}
	\begin{tabular*}{\textwidth}{@{\extracolsep{4pt}} lrlrrr}
		\toprule
		\addlinespace[1em]
		Category & \multicolumn{2}{c}{Mean unit price} & \# Vendors & \# Offers & \multicolumn{1}{c}{Median quantity} \\ 
		\midrule
		Cannabis & $10.64$ & per 1g & $1,314$ & $5,902$ & $14.00$ g \\ 
		MDMA & $46.00$ & per 1g & $721$ & $3,368$ & $7.00$ g \\ 
		Cocaine & 109.19 & per 1g & $679$ & $2,898$ & $3.50$ g \\ 
		Amphetamine & $15.88$ & per 1g & $319$ & $1,469$ & $20.00$ g \\
		LSD & $4.81$ & per $10 \times 100 \mu$g & $223$  & $1,465$ & $31.25 \times 100 \mu$g \\
		Meth & $158.15$ & per 1g & $296$ & $1,247$ & $3.50$ g \\
		Heroin & $152.00$ & per 1g & $244$ & $1,054$ & $1.00$ g \\ 
		Ketamine & $58.46$ & per 1g & $127$ & $531$ & $5.00$ g \\
		\bottomrule
	\end{tabular*}
	\fnote{\textbf{Notes:} The table reports summary statistics for the eight product categories considered. Prices are reported in USD.\@ The Bitcoin exchange rate used corresponds to the day on which the item price was observed. Unit price refers to the price per consumption unit, defined as 1 gram for all categories except LSD, where it is 100 $\mu$g.}
\end{table}

\begin{table} \centering
  \captionsetup{width=\textwidth,justification=raggedright,singlelinecheck=off}
  \caption{Re-entering Evolution sellers before and after the exit}
  \label{table:sw-prevspost}
  \small
  \begin{tabular*}{\textwidth}{@{\extracolsep{5pt}}lccccccccc}
    \toprule
    \addlinespace[1em]
    & \multicolumn{2}{c}{Mean \# offers} & \multicolumn{2}{c}{Median quantity} &  \multicolumn{4}{c}{Mean unit price}  \\
    \cmidrule{2-3} \cmidrule{4-5} \cmidrule(r){6-9}
    Category	&	Before	&	After	&	Before	&	After	&	Before	&	After	&	Diff		&	$p$-value	\\ 
    \hline \\[-1.8ex]
    All			&	$2.87$	&	$2.78$	&	$7$		&	$7$		&	$42.72$ &	$36.71$ &	$$-$6.01$	&	$0.07$	\\ 
    Cannabis	&	$2.17$	&	$2.09$	&	$10$	&	$10$	&	$10.75$ &	$10.10$ &	$$-$0.65$	&	$0.10$	\\ 
      Cocaine		&	$2.68$	&	$2.38$	&	$3$		&	$3.50$	&	$86.09$ &	$79.48$ &	$$-$6.62$	&	$0.08$	\\ 
      Amphetamine &	$2.71$	&	$3.80$	&	$25$	&	$25$	&	$8.61$	&	$7.77$	&	$$-$0.84$	&	$0.07$	\\ 
    \bottomrule
  \end{tabular*}
  \fnote{\textbf{Notes:} The table contrasts re-entering Evolution sellers one week prior to the exit and two weeks after the exit. It shows across categories i) the average number of offers per seller at the two dates, ii) the median quantity of offers at the two dates, iii) the absolute difference in mean unit prices charged by Evolution sellers and by all other sellers, and the p-value of a t-test of the difference of price changes. Prices are reported in USD.\@ The Bitcoin exchange rate used corresponds to the day on which the item price was observed. Unit price refers to the price per consumption unit, defined as 1 gram for all categories listed in the table.}
\end{table}

\begin{table}
\centering
\captionsetup{width=\textwidth,justification=raggedright,singlelinecheck=off}
\caption{The effect of re-entry on rating}
\label{tab:fs}
\small
\begin{threeparttable}
\begin{tabular*}{\textwidth}{@{\extracolsep{\fill}} l*{5}{S[%
	table-number-alignment=center,
	table-figures-integer=2, table-figures-decimal=3,
	round-precision=3,
	table-space-text-pre={-},
	table-space-text-post={***}
			]}}
\toprule
\addlinespace[2em]
	& {All} & {Cannabis} & {Cocaine} & {Amphetamine} \\
	\midrule
	Re-entry x post         & -0.089{**} & -0.092  & -0.092  & -0.080  \\
							& (0.042)     & (0.065) & (0.075) & (0.080) \\
							& [0.035]	 & [0.160]	& [0.220] & [0.316] \\
	Escrow                  & 0.018{*}   & 0.018   & 0.021   & 0.011   \\
							& (0.011)     & (0.015) & (0.019) & (0.013) \\
							& [0.099]	 & [0.235]	& [0.281] & [0.375] \\
	\midrule
	Item-specific FE        & {\checkmark} & {\checkmark} & {\checkmark} & {\checkmark} \\	
	Time FE        			& {\checkmark} & {\checkmark} & {\checkmark} & {\checkmark} \\		
	Num. obs.               & {50254}	   & {27220}   & {14947}   & {8087}    \\
	Num. items\    			& {6656}        & {3708}    & {1854}    & {1094}    \\
	\bottomrule
\end{tabular*}
    \begin{tablenotes}[para,flushleft]\scriptsize
      \item \textbf{Notes:} Results based on a linear model as specified in \autoref{sec:Approach} with weekly data. The sample is restricted to a time period around the evolution exit, from 15 weeks prior to 10 weeks after the exit. Standard errors clustered at seller-substance level given in parentheses, $p$-values in brackets. *, ** and *** denote p$<$0.1, p$<$0.05 and p$<$0.01, respectively.
    \end{tablenotes}
  \end{threeparttable}
\end{table}


\begin{table}
\centering
\captionsetup{width=\textwidth,justification=raggedright,singlelinecheck=off}
\caption{The effect of rating on price}
\small
\begin{threeparttable}
	\begin{tabular*}{\textwidth}{@{\extracolsep{\fill}} l*{5}{S[%
	table-number-alignment=center,
	table-figures-integer=2, table-figures-decimal=3,
	round-precision=3,
	table-space-text-pre={-},
	table-space-text-post={***}
	]}}
	\toprule
	\addlinespace[2em]
	 & {All} & {Cannabis} & {Cocaine} & {Amphetamine} \\
	\midrule
	Rating (ihs)            & 1.036{*} & 0.750   & 0.505   & 2.251   \\
							& (0.550)   & (0.593) & (0.576) & (2.363) \\
							& [0.059]	& [0.21]  & [0.38]  & [0.34] \\
	Escrow                  & -0.000    & -0.001  & 0.017   & -0.002  \\
							& (0.011)   & (0.013) & (0.014) & (0.028) \\
							& [0.967]	& [0.947] & [0.225] & [0.950] \\
	\midrule
	Item-specific FE        & {\checkmark} & {\checkmark} & {\checkmark} & {\checkmark} \\	
	Time FE        			& {\checkmark} & {\checkmark} & {\checkmark} & {\checkmark} \\		
	Num. obs.               & {50254}     & {27220}   & {14947}   & {8087}    \\
	Num. items     & {6656}      & {3708}    & {1854}    & {1094}    \\
	\bottomrule
	\end{tabular*}
	\begin{tablenotes}[para,flushleft]\scriptsize
	  \item \textbf{Notes:} Results based on a linear model as specified in \autoref{sec:Approach} with weekly data. The sample is restricted to a time period around the evolution exit, from 15 weeks prior to 10 weeks after the exit. Standard errors clustered at item-level given in parentheses, $p$-values in brackets. *, ** and *** denote p$<$0.1, p$<$0.05 and p$<$0.01, respectively.
	\end{tablenotes}
\end{threeparttable}
\label{tab:iv}
\end{table}


\clearpage

\section{Appendix: Figures and Tables}
\setcounter{table}{0}
\setcounter{figure}{0}
\renewcommand\thetable{\Alph{section}.\arabic{table}}
\renewcommand\thefigure{\Alph{section}.\arabic{figure}}
\label{sec:Appendix}

\begin{figure}[h]
	\caption{Screenshots from Agora}\vspace{-0.5em}
	\label{fig:agoscreens}
	\begin{subfigure}[b]{0.5\linewidth}
		\centering
		\includegraphics[width=0.97\linewidth,height=0.35\textheight]{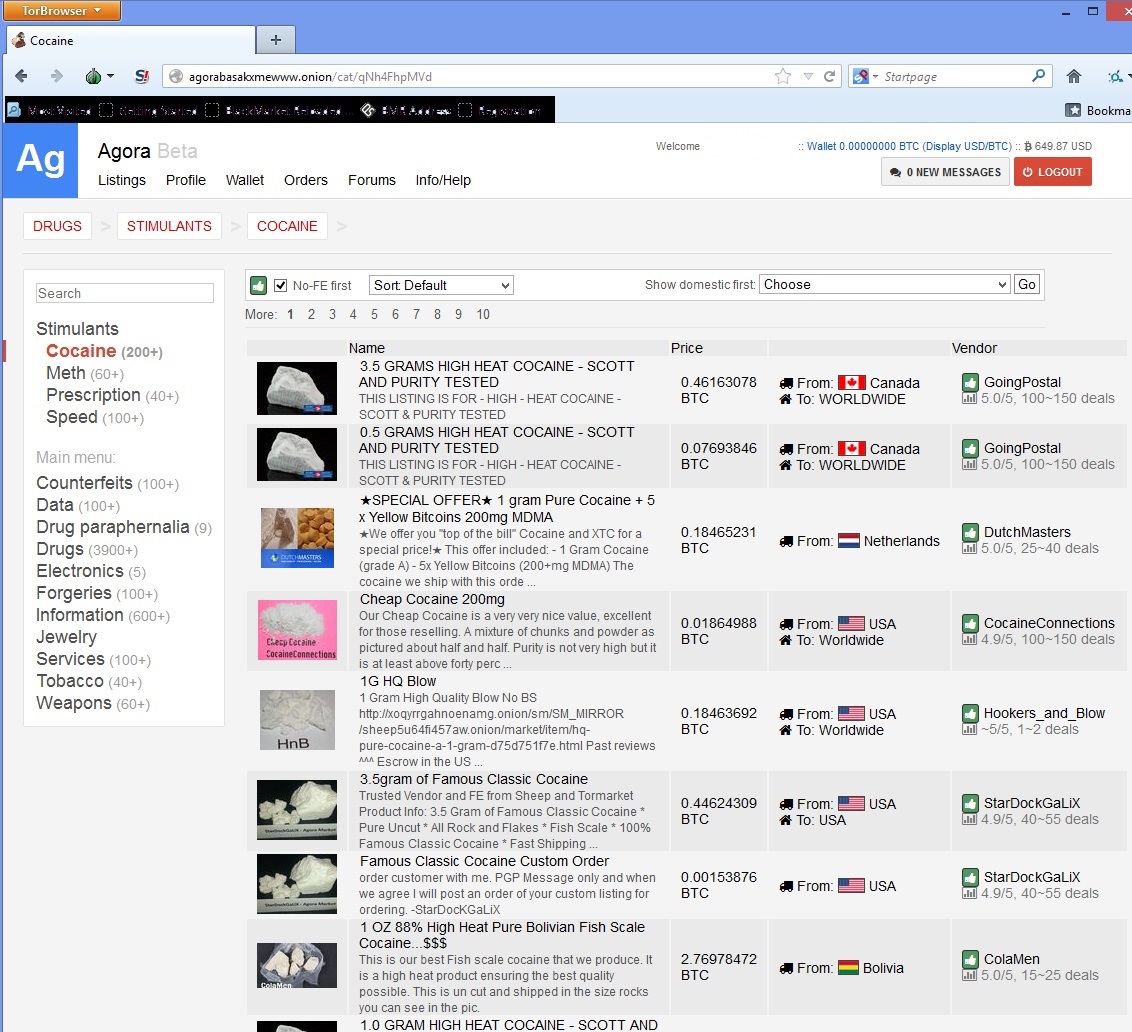}
	\end{subfigure}
	\begin{subfigure}[b]{0.5\linewidth}
		\centering
		\includegraphics[width=0.97\linewidth,height=0.35\textheight]{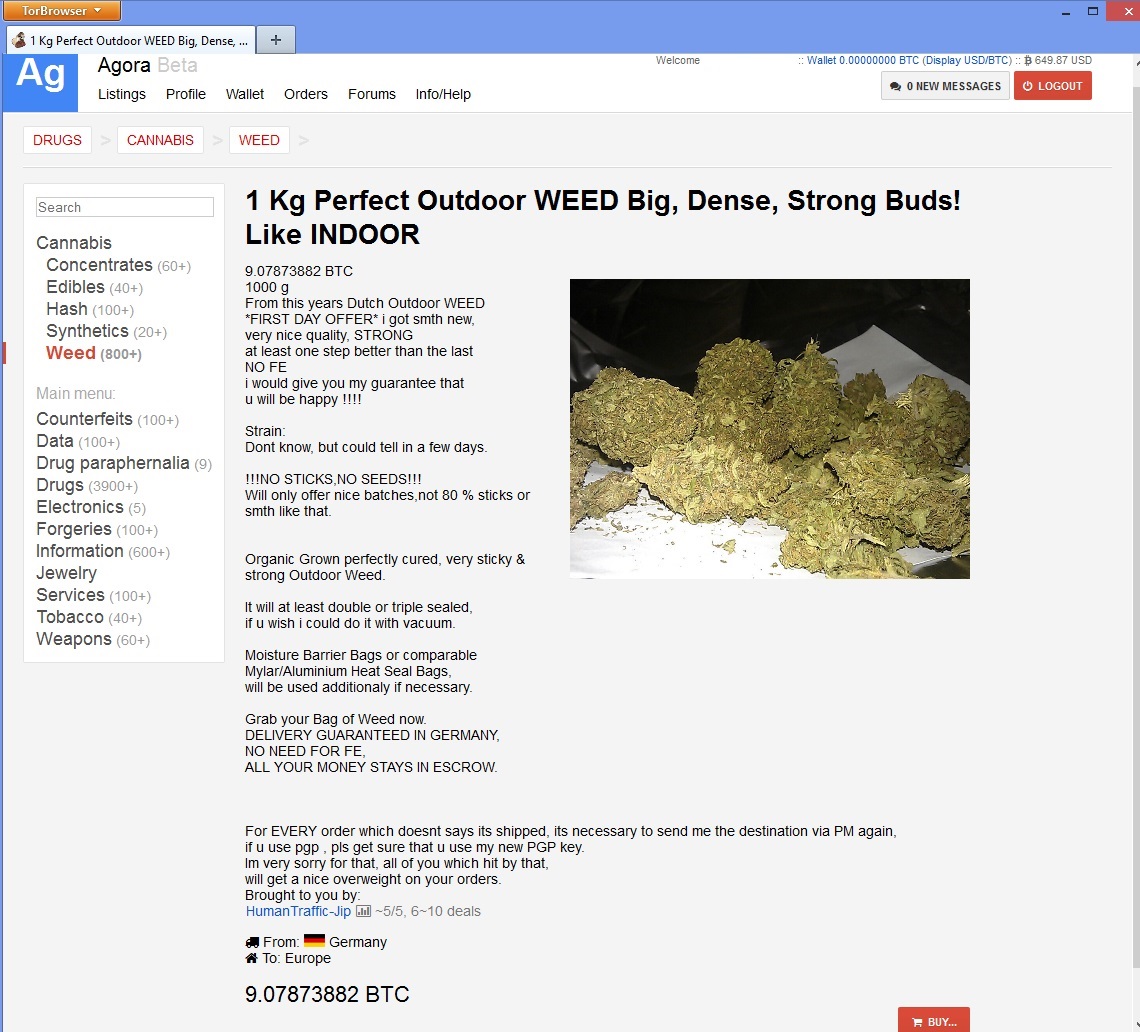}
	\end{subfigure}
	\fnote{\textbf{Notes:} The figure shows two screenshots of the darknet platform Agora as it appears to potential buyers browsing.}
\end{figure}

\begin{figure}
	\centering
	\caption{Screenshot of Evolution item}\vspace{-0.5em}
	\label{fig:evo-lsdoffer}
	\includegraphics[width=0.98\linewidth]{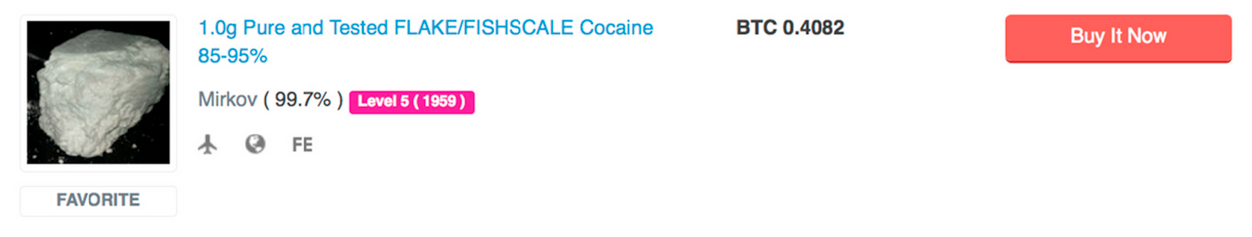}
	\fnote{\textbf{Notes:} The figure shows a screenshot of of an item being sold on the darknet platform Evolution.}
\end{figure}

\begin{figure}
	\caption{Screenshot of the Grams search engine website}
	\label{fig:grams}
	\centering
	\includegraphics[width=0.95\linewidth]{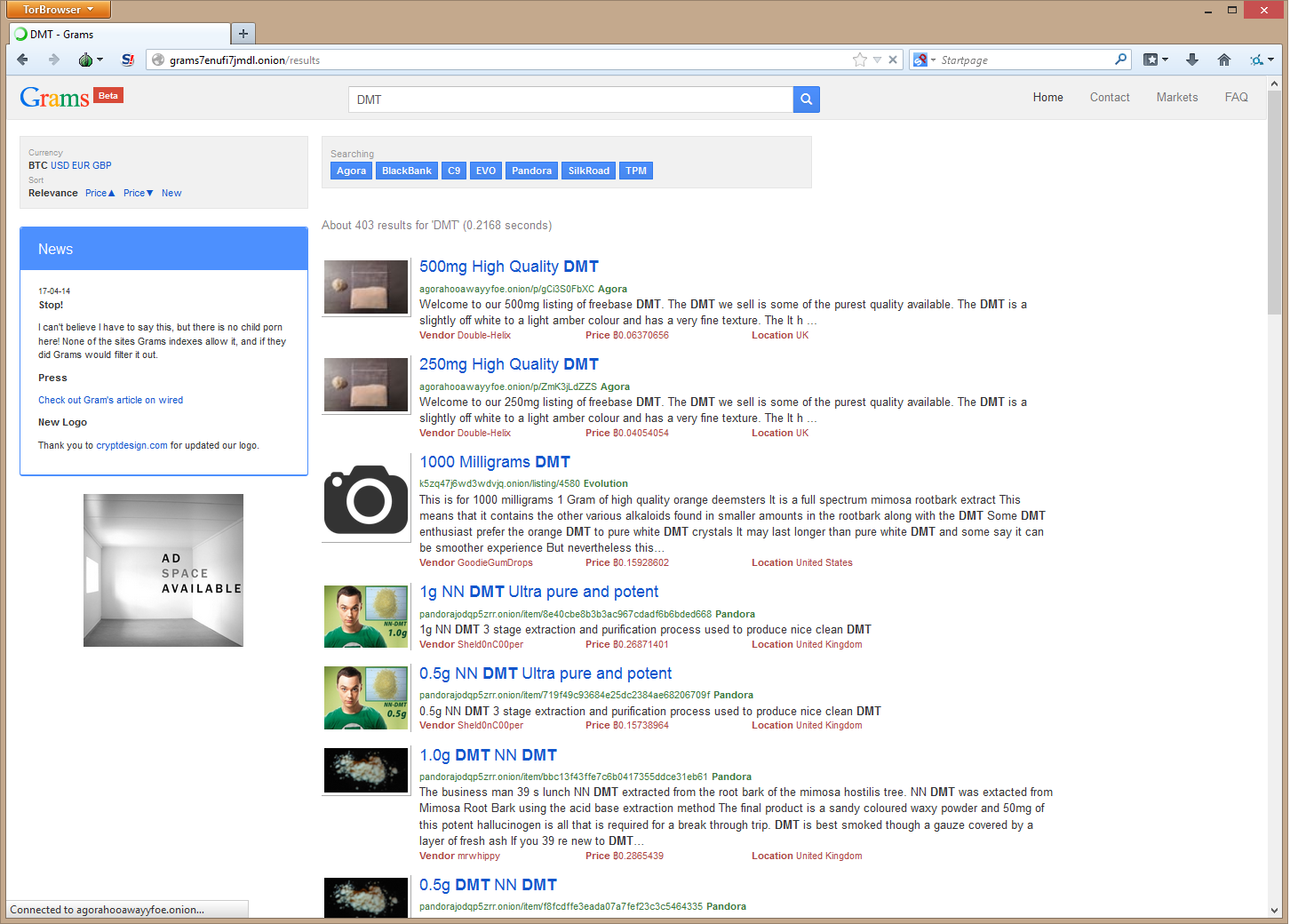}
\end{figure}

\begin{figure}
	\caption{Silk Roads payment system}
	\label{fig:escrow}
	\centering
	\includegraphics[width=0.95\linewidth]{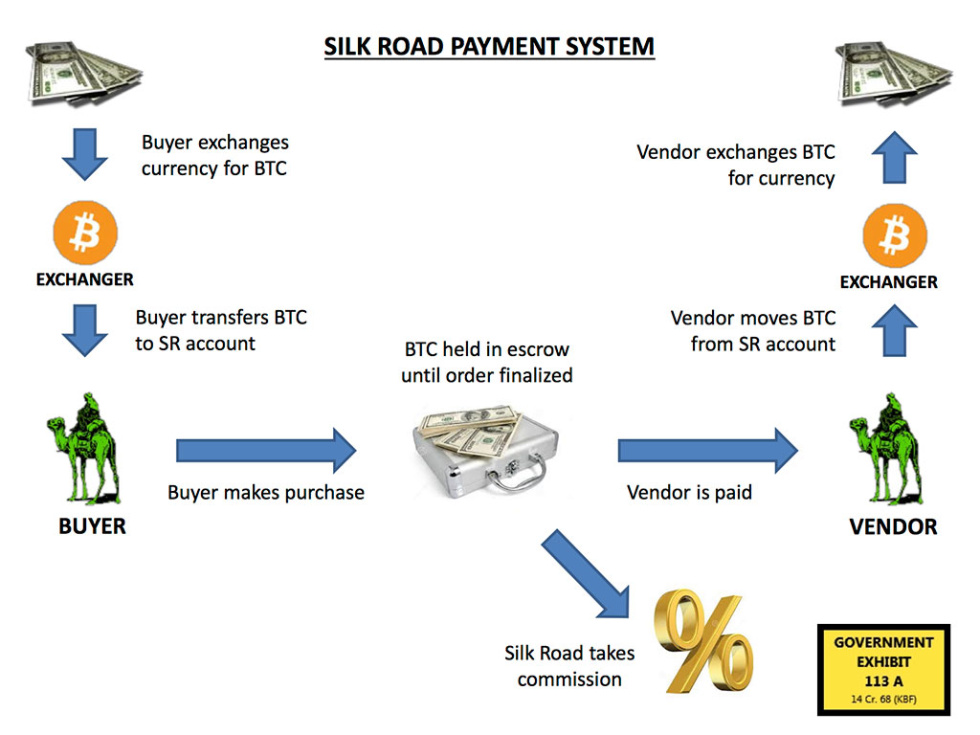}
	\fnote{\textbf{Notes:} The figure shows the payment system originated by Silk Road. Using the escrow system of the platform, buyers may transfer payment onto the escrow account instead of sending directly to the seller. Finalizing the order refers to buyers signalling receipt of the goods. Source: US government diagram used in the Silk Road trial, arstechnica.com.}
\end{figure}



\begin{figure}
	\centering
	\caption{Distributions of quantity and unit price by category of drug}
	\label{fig:histquantppc}
	\begin{subfigure}{\textwidth}
		\caption{Cannabis}
		\includegraphics[height = 0.29\textheight]{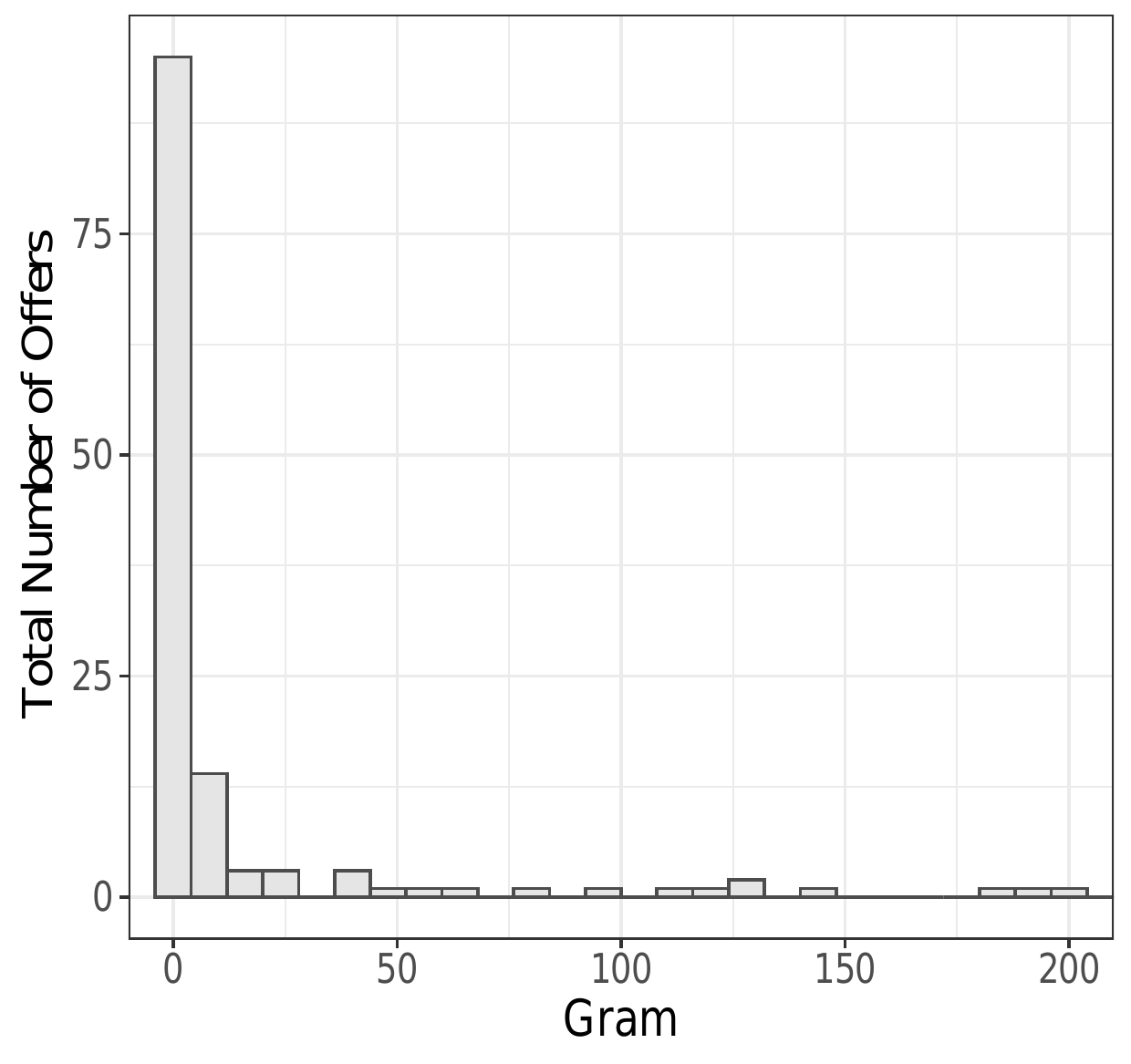}
		\hfill
		\includegraphics[height = 0.29\textheight]{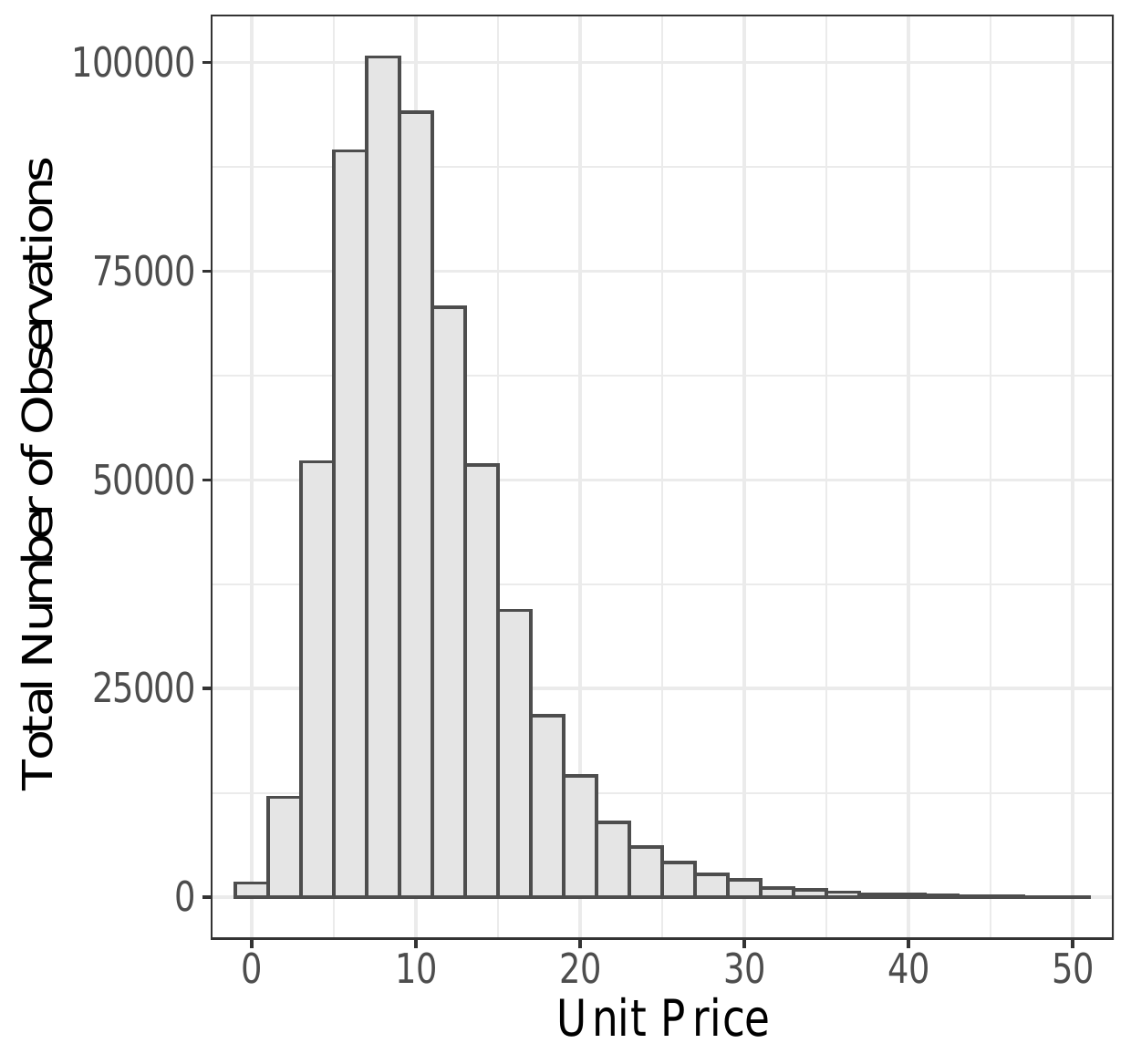}
	\end{subfigure}
	\vspace{1em}
	\begin{subfigure}{\textwidth}
	\caption{Cocaine}
	\includegraphics[height = 0.29\textheight]{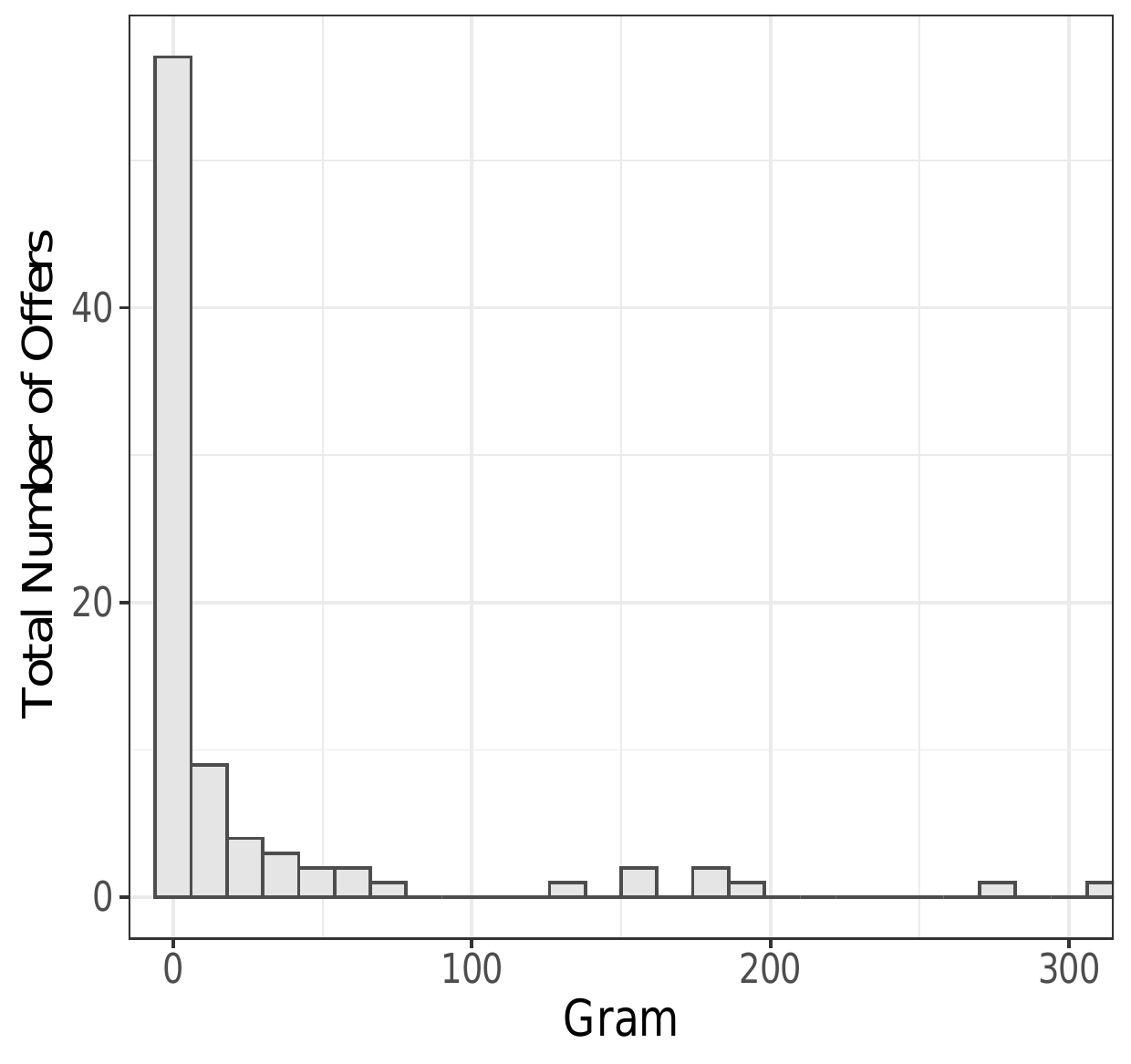}
	\hfill
	\includegraphics[height = 0.29\textheight]{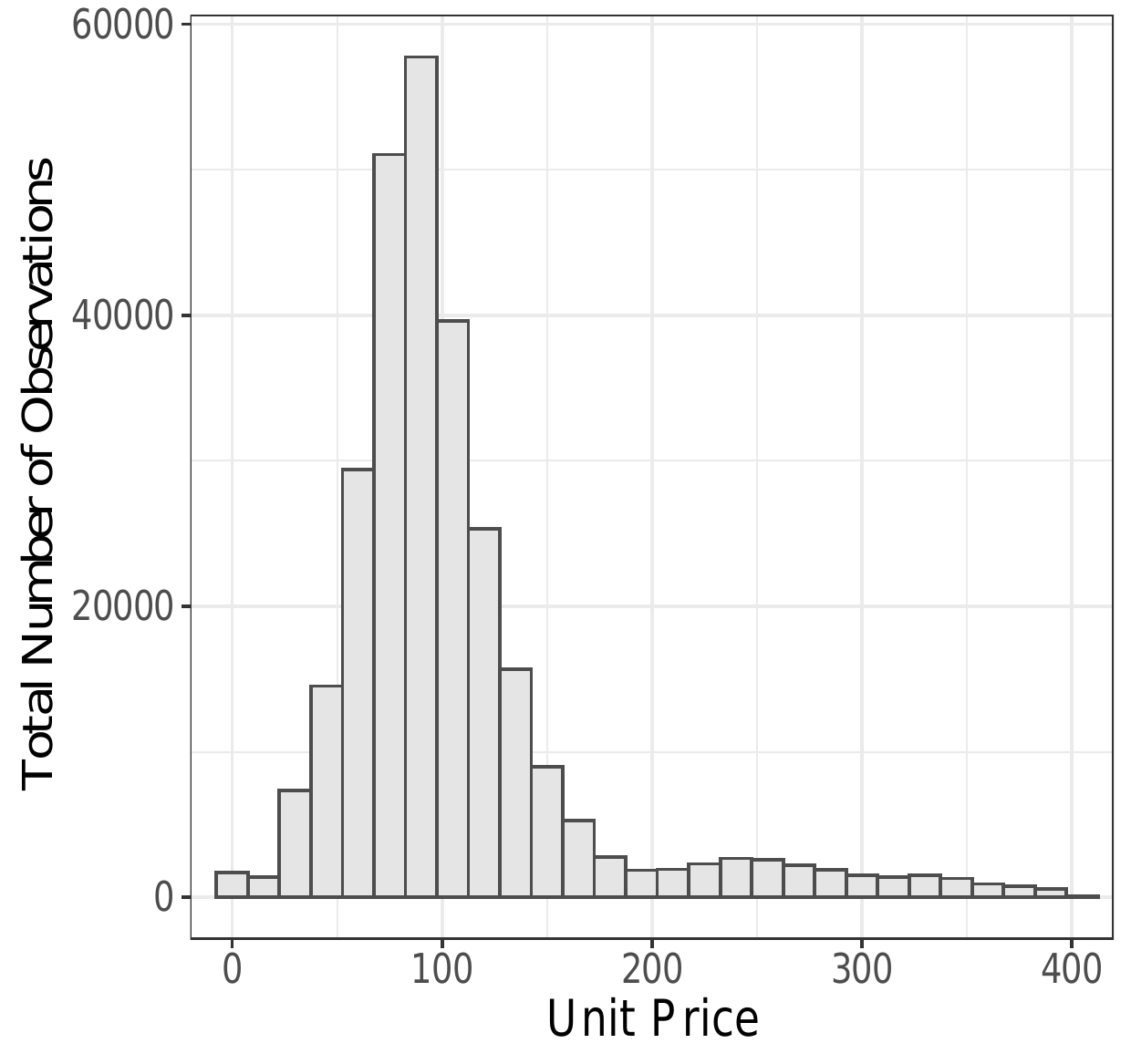}
	\end{subfigure}
	\vspace{1em}
	\begin{subfigure}{\textwidth}
	\caption{Amphetamine}
	\includegraphics[height = 0.29\textheight]{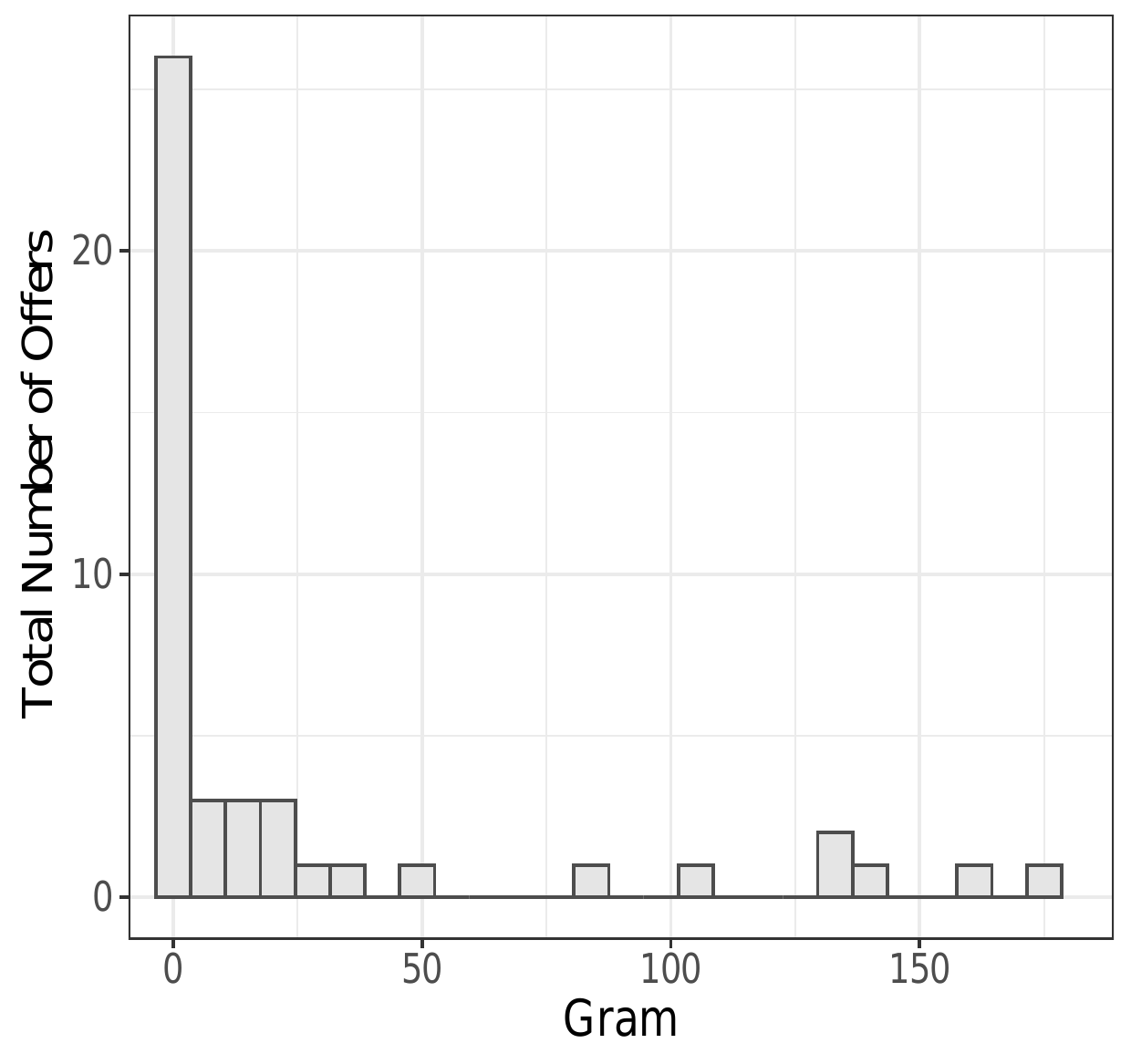}
	\hfill
	\includegraphics[height = 0.29\textheight]{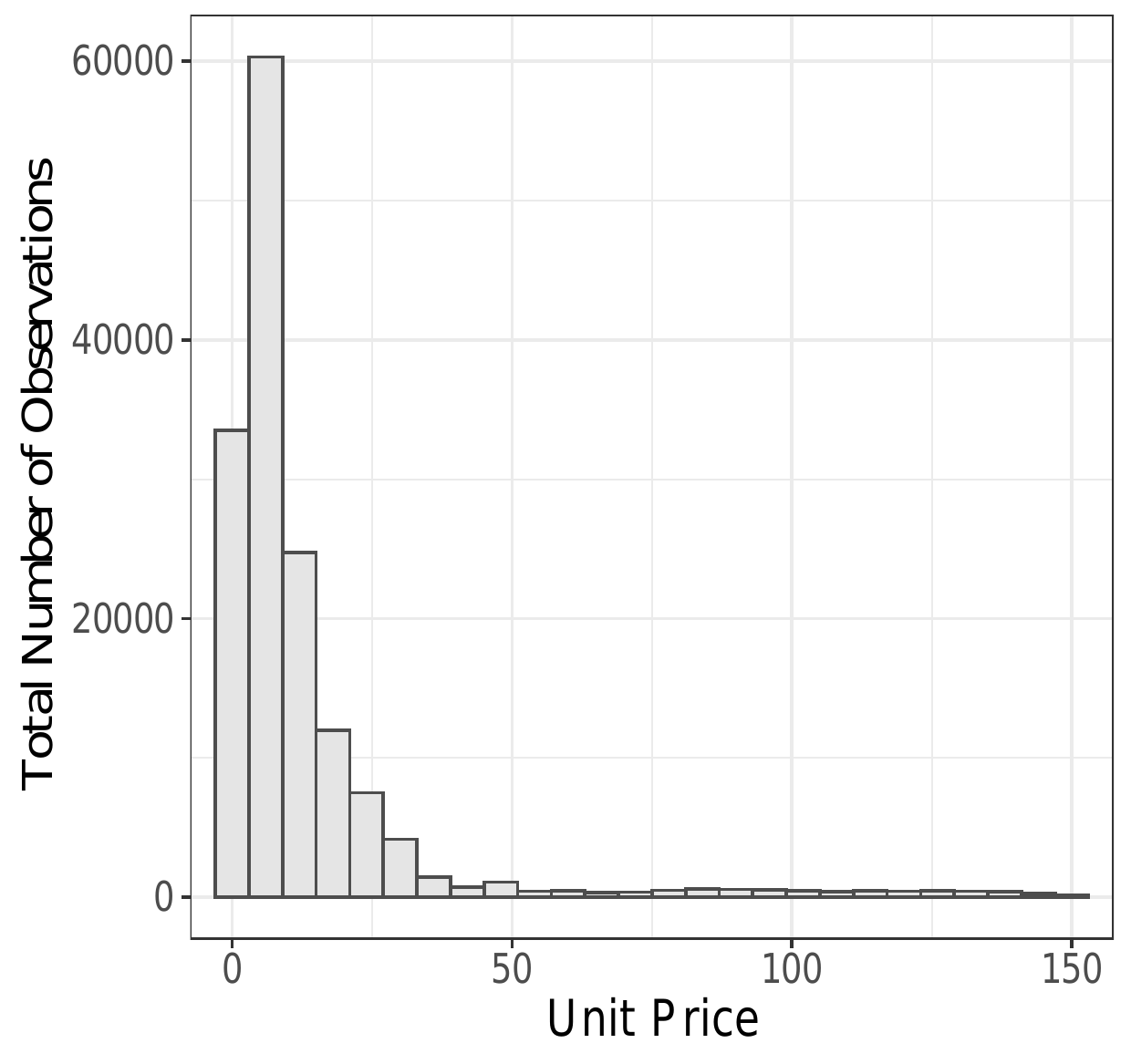}
	\end{subfigure}
\end{figure}

\begin{figure}
	\caption{Number of unique offers for illegal drugs by shipping origin}
	\label{fig:map}
	\centering
	\includegraphics[width=\linewidth]{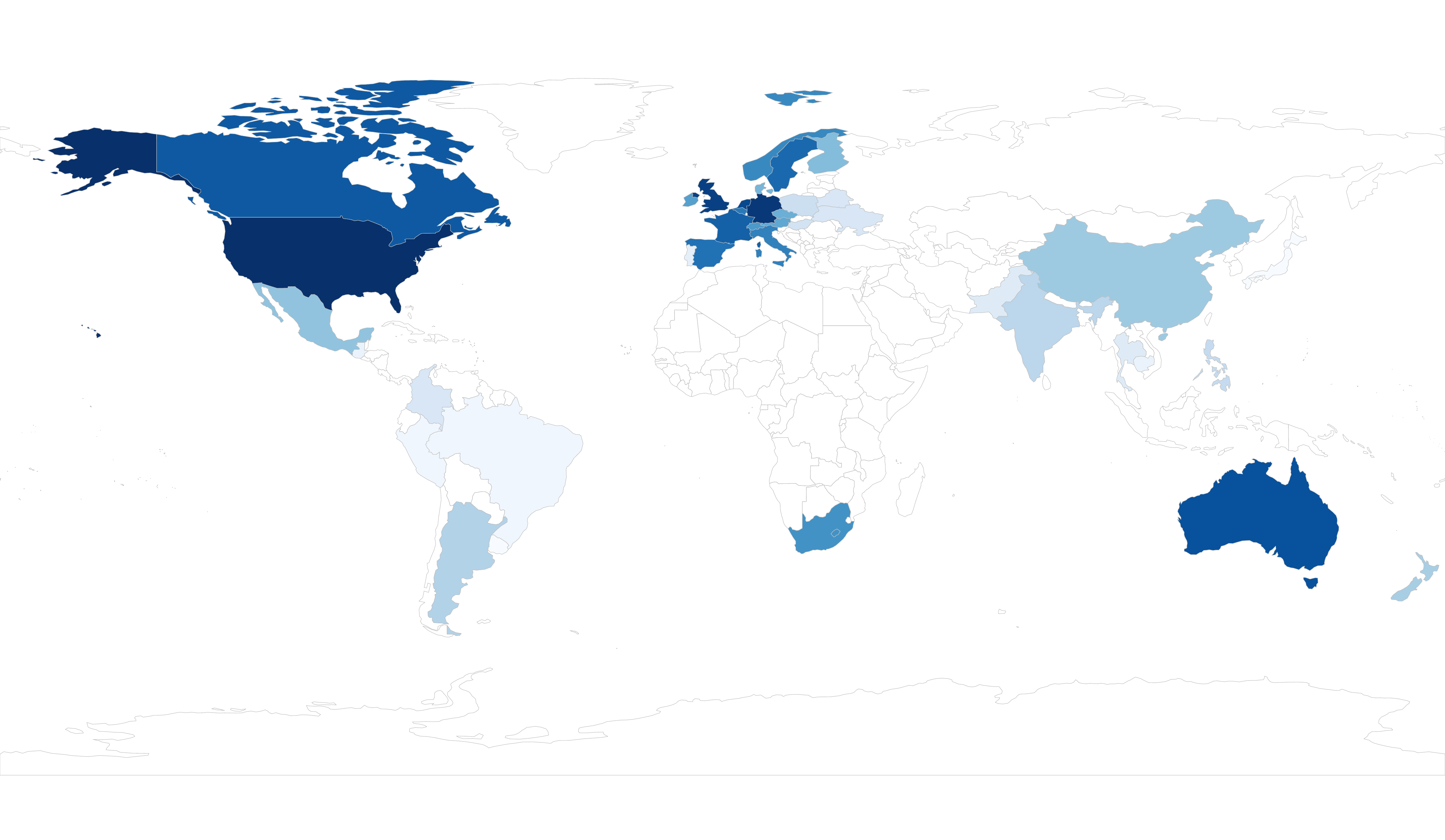}
	\fnote{\textbf{Notes:} The figure shows the total number of unique items shipped from each country on both platforms. The largest market is the United States. Most of the offers originate in North America, (Western) Europe, and Australasia.}
\end{figure}

\begin{figure}
	\centering
	\caption{Platform uptime}
	\label{fig:uptime}
	\includegraphics[width=\linewidth]{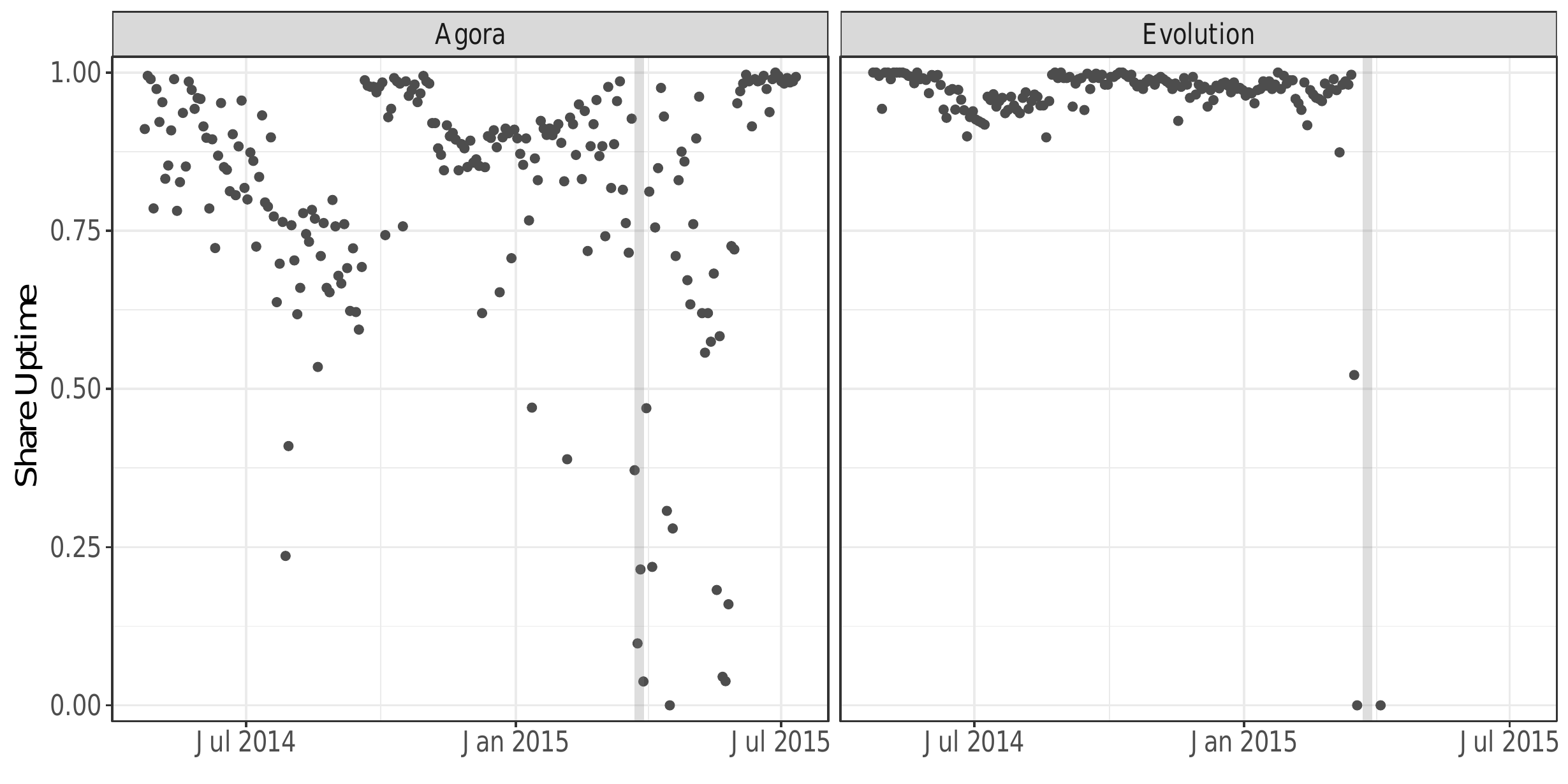}
	\fnote{\textbf{Notes:} The figure shows the percentage share of uptime for each of the two platforms. The Evolution exit is indicated in grey.}
\end{figure}

\begin{figure}
	\centering
	\caption{Distribution of seller rating by platform}
	\label{fig:rathist}
	\begin{subfigure}{0.49\textwidth}
		\caption{Agora}
		\includegraphics[width=\linewidth]{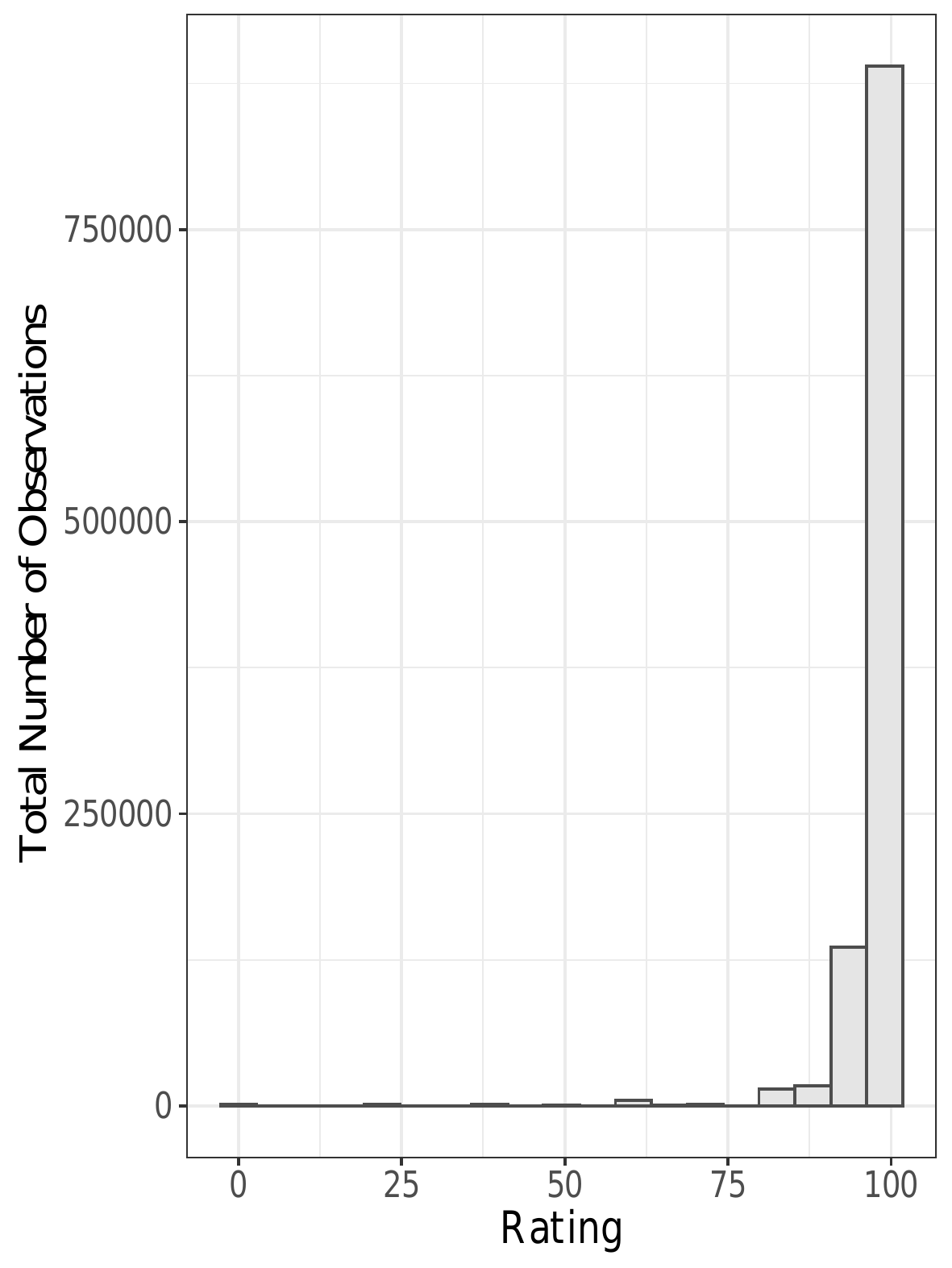}
	\end{subfigure}
		\hfill
	\begin{subfigure}{0.49\textwidth}
		\caption{Evolution}
		\includegraphics[width=\linewidth]{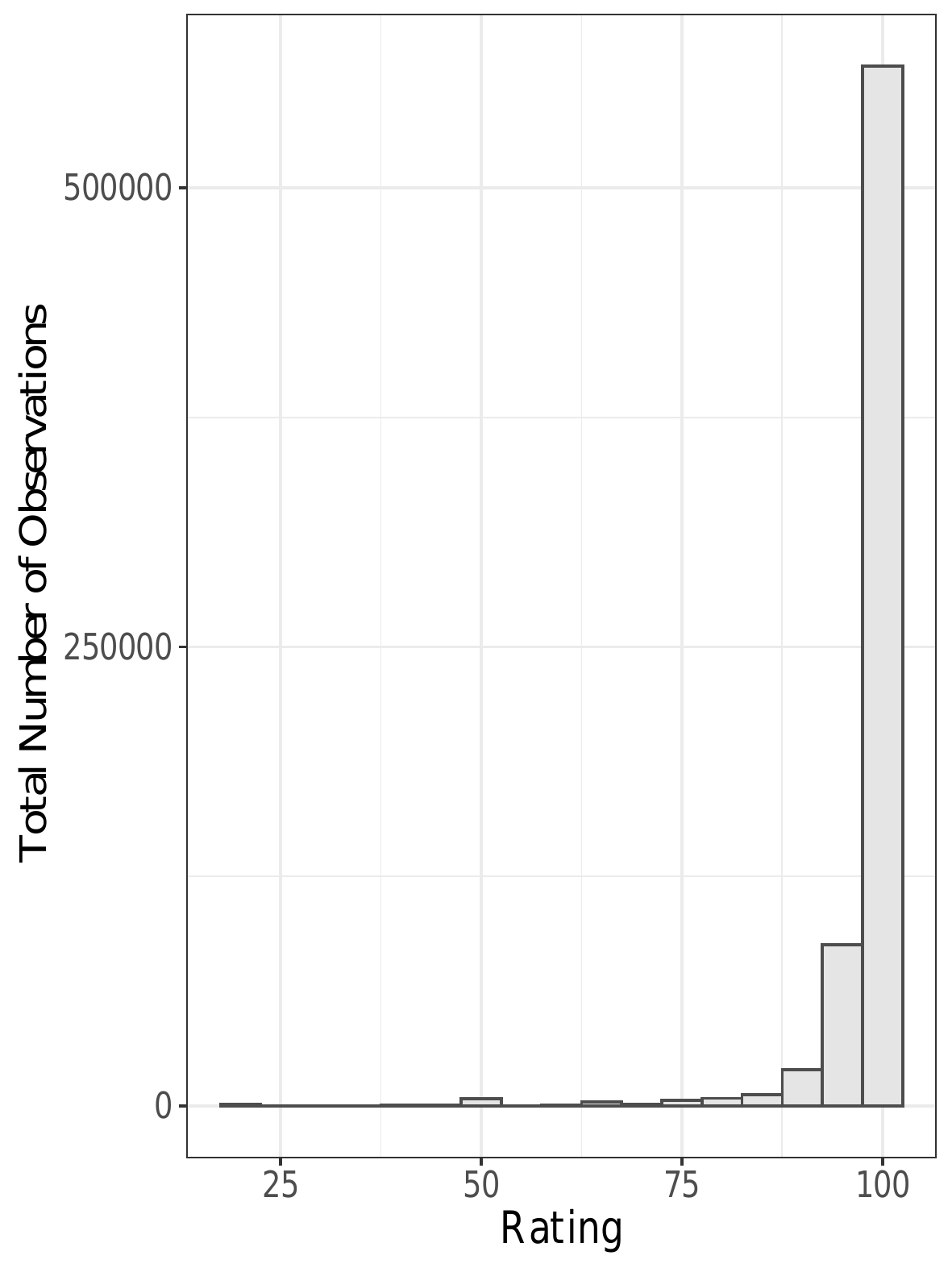}
	\end{subfigure}
	\fnote{\textbf{Notes:} The figure shows the distribution of ratings of the individual vendor accounts. The rating is defined from 0 to 100, where a higher number indicates a better rating.}
\end{figure}

\begin{figure}
	\caption{An example of a PGP key}
	\label{fig:pgpkey}
	\centering
	\includegraphics[width=0.85\linewidth]{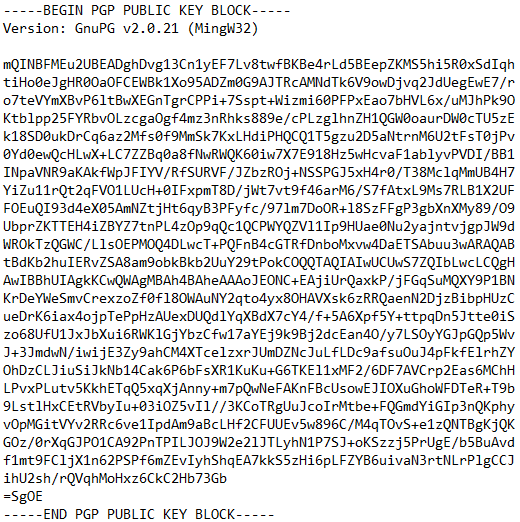}
	\vspace{1em}
	\fnote{\textbf{Notes:} The figure shows an example of a public PGP key block. The key can be used to encrypt information sent to the owner of the private PGP key. These PGP key blocks are provided by the sellers on their account information and directly visible to buyers.}
\end{figure}

\begin{figure}
	\caption{The bitcoin exchange rate}
	\label{fig:btcxr}
	\centering
	\includegraphics[width=\linewidth]{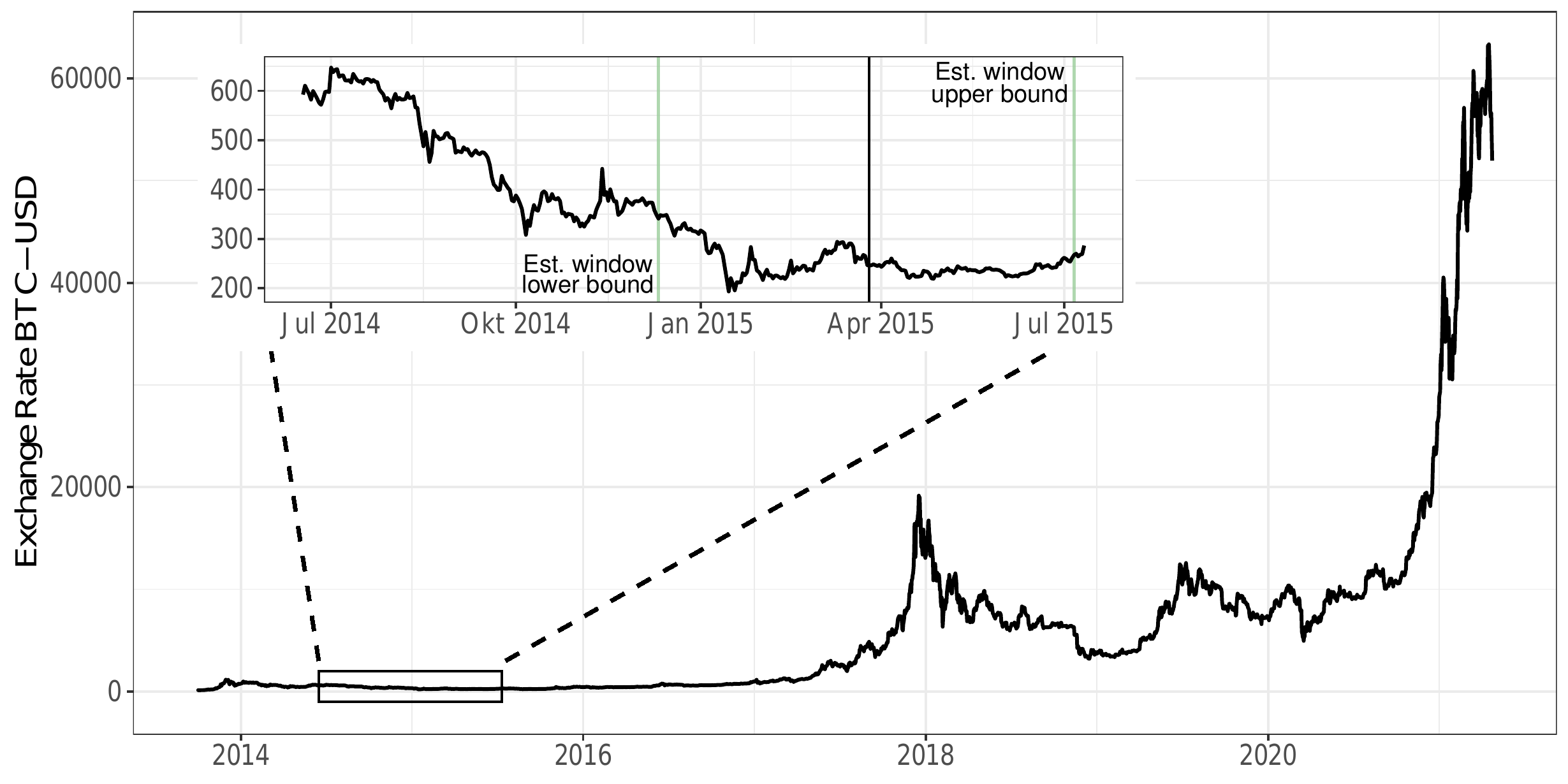}
	\fnote{\textbf{Note:} The figure depicts the Bitcoin-USD exchange rate from 2014 to 2021. The highlighted segment shows the exchange rate in the timeframe studied in this paper. The Evolution exit is indicated by the vertical black line.}
\end{figure}

\begin{figure}
	\centering
	\caption{Share of re-entering sellers during the month preceding exit (top) and during the month following exit (bottom)}\vspace{-0.5em}
	\label{fig:sw-histdrop}
	\includegraphics[width=\linewidth]{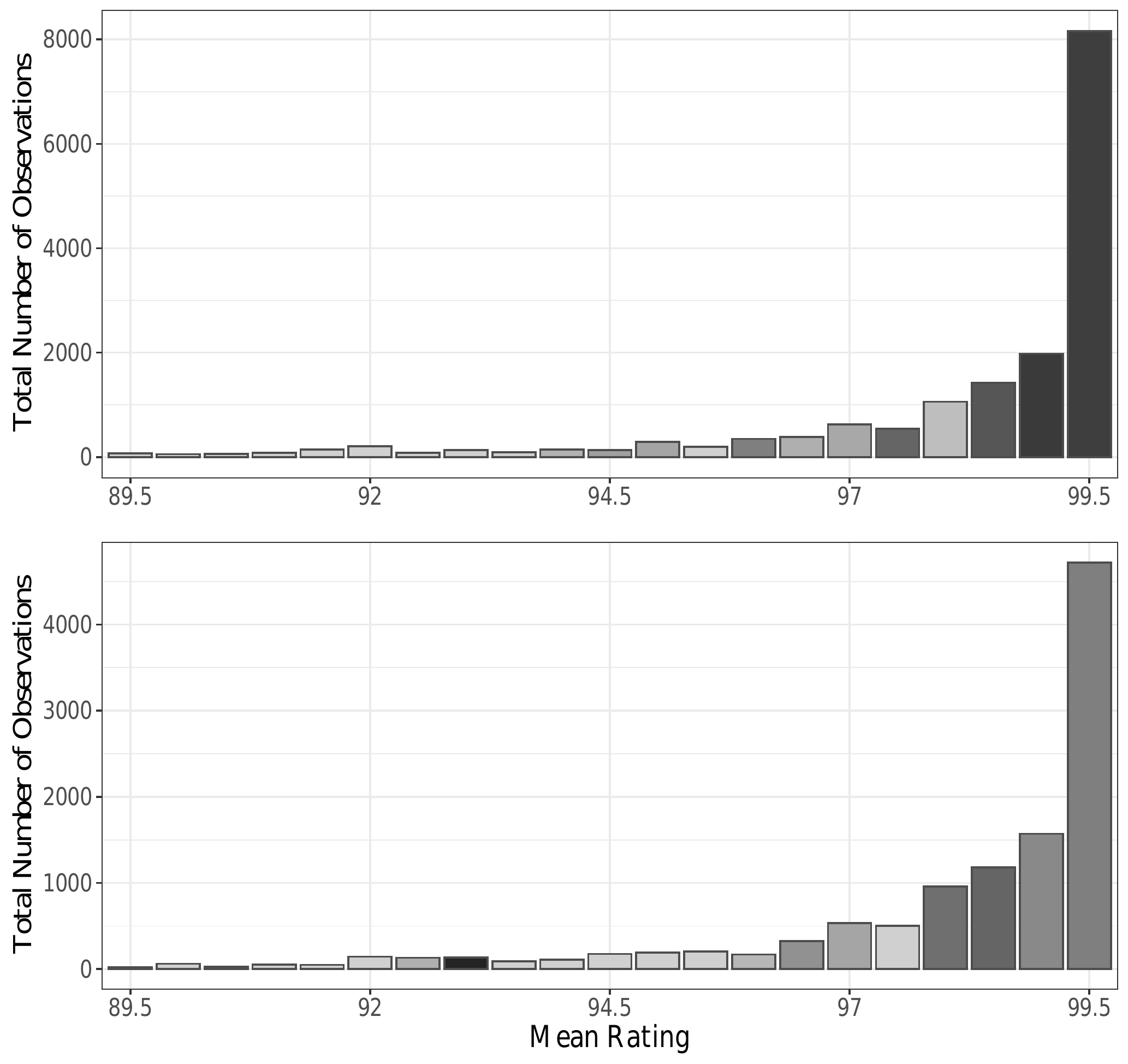}
	\fnote{\textbf{Notes:} The figure shows the top end of the distribution of seller rating during the 28 days preceding the Evolution exit (top) and the 28 days following the exit (bottom). The shading corresponds to the share of re-entering sellers in each bar, with darker coloring indicating a higher share of re-entering sellers.}
\end{figure}



 \begin{figure}
 	\centering
 	\caption{Evolution of feedback count over time}
 	\label{fig:shock-fb}
 	\includegraphics[width=\linewidth]{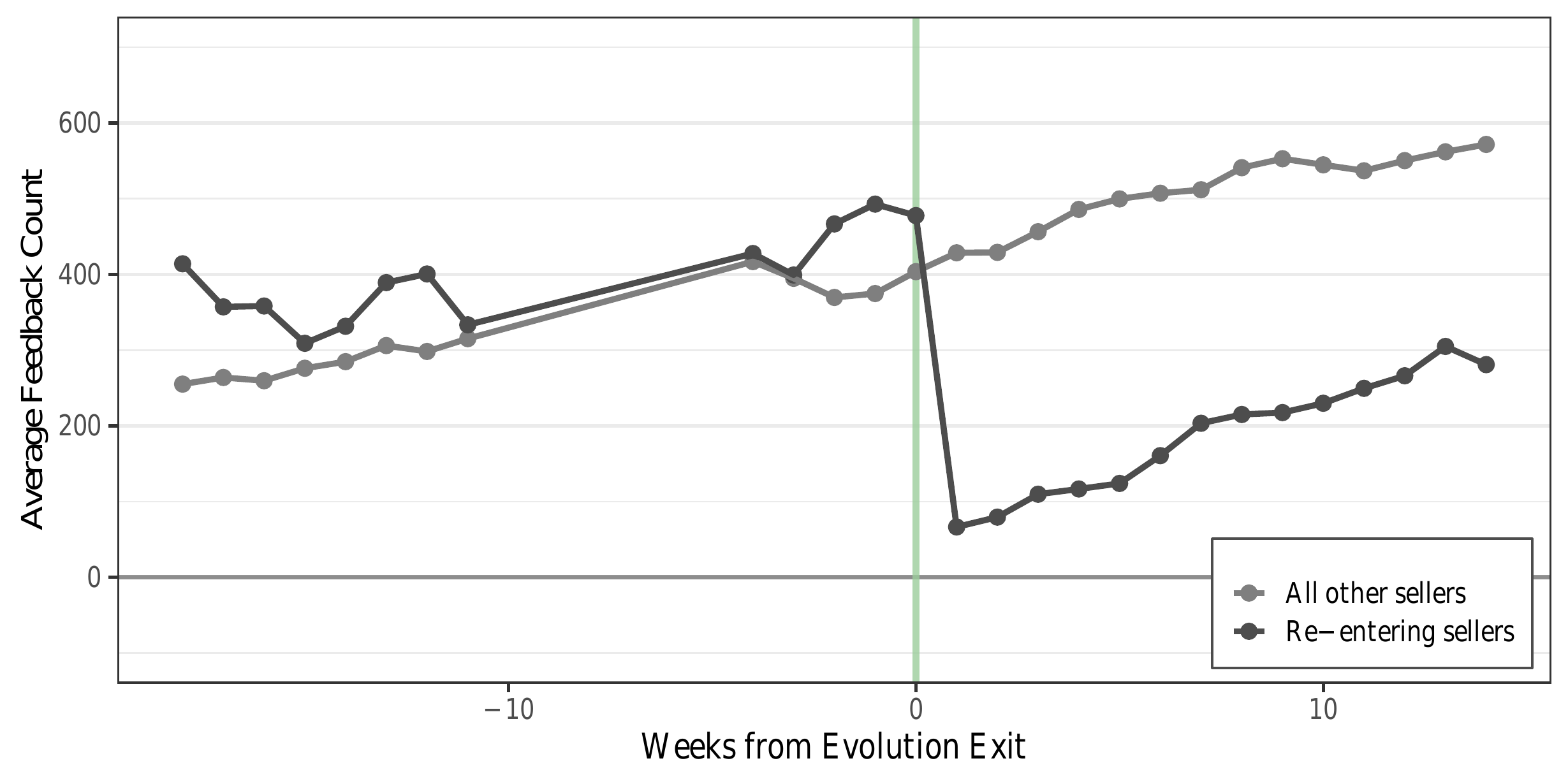}
 	\fnote{\textbf{Notes:} The figure shows the mean feedback count of re-entering sellers (sellers that sold exclusively on Evolution before the exit and migrated to Agora following the exit) and all other sellers both before and after the exit. The Evolution exit period is highlighted in green.}
 \end{figure}



\clearpage



\begin{table}
	\captionsetup{width=\textwidth,justification=raggedright,singlelinecheck=off}
	\centering
	\caption{Platform comparison}
	\label{tab:platf-diff}
	\small
	\begin{tabular*}{\textwidth}{@{\extracolsep{5pt}} lcccccccc}
		\toprule
		\addlinespace[1em]
		& \multicolumn{2}{c}{Mean Price} & \multicolumn{2}{c}{Median Quantity} & \multicolumn{2}{c}{\# Sellers} &  \multicolumn{2}{c}{\# Items} \\
		\cmidrule{2-3} \cmidrule{4-5} \cmidrule{6-7} \cmidrule{8-9}
		Category & Agora & Evol.\ & Agora & Evol.\ & Agora & Evol.\ & Agora & Evol.\ \\
		\midrule
		Cannabis & 10.34 & 10.51 & 20.00 & 10.00 & 554 & 589 & 2195 & 2326 \\
		Cocaine & 93.38 & 93.95 & 3.50 & 3.50 & 258 & 302 & 1005 & 1207 \\
		Amphetamine & 9.99 & 10.02 & 20.00 & 20.00 &  99 & 173 & 370 & 789 \\
		\bottomrule
	\end{tabular*}
	\fnote{\textbf{Notes:} The table shows the mean unit price, median quantity, number of sellers and number of items observed on either platform seven days prior to the Evolution exit. Prices are reported in USD\@. The Bitcoin exchange rate used corresponds to the day on which the item price was observed. Unit price refers to the price per 1 gram.}
\end{table}

\begin{table} \centering
	\captionsetup{width=\textwidth,justification=raggedright,singlelinecheck=off}
	\caption{Country variation for cocaine}
	\label{table:countrydiffcoc}
	\small
	\begin{tabular*}{\textwidth}{@{\extracolsep{5pt}} lcccc}
		\toprule
		\addlinespace[1em]
		Shipping origin country & Mean unit price & \# Offers & \# Vendors & \multicolumn{1}{c}{Median quantity} \\
		\hline \\[-1.8ex]
		United States & 81.10 & 753 & 186 & 3.50 \\ 
		Netherlands & 86.32 & 493 &  92 & 3.50 \\ 
		United Kingdom & 110.58 & 461 & 108 & 3.00 \\ 
		Germany & 93.52 & 276 &  57 & 5.00 \\ 
		Australia & 283.32 & 272 &  77 & 2.00 \\ 
		Canada & 94.56 & 201 &  41 & 3.50 \\ 
		France & 105.54 &  74 &  18 & 1.00 \\ 
		Belgium & 88.10 &  44 &   9 & 5.00 \\ 
		Sweden  & 128.36 &  41 &  12 & 2.00 \\ 
		Italy & 101.43 &  32 &   5 & 5.00 \\ 
		\bottomrule
	\end{tabular*}
	\fnote{\textbf{Notes:} The table reports summary statistics for cocaine for the ten largest countries of origin as measured by the number of vendors active, sorted by size. Prices are reported in USD. The Bitcoin exchange rate used corresponds to the day on which the item price was observed. Unit price refers to the price per 1 gram.}
\end{table}

\begin{table}
	\captionsetup{width=\textwidth,justification=raggedright,singlelinecheck=off}
	\caption{Quantity discounts}
	\label{table:discounts}
	\small
	\begin{tabular*}{\textwidth}{@{\extracolsep{20pt}} lccccc}
		\toprule \addlinespace[1em]
		& & \multicolumn{4}{c}{Discounts}\\
		\cmidrule(rl){3-6}
		Category &  Single Unit & $\times$ 5 & $\times$ 10 & $\times$ 50 & $\times$ 100 \\
		\hline \\[-1.8ex]
		Cannabis & 17.53 & 0.77 & 0.67 & 0.49 & 0.42 \\ 
		MDMA & 65.97 & 0.63 & 0.51 & 0.33 & 0.27 \\ 
		Cocaine & 125.38 & 0.74 & 0.66 & 0.57 & 0.53 \\ 
		Amphetamine & 42.54 & 0.37 & 0.25 & 0.14 & 0.10 \\ 
		Meth & 172.79 & 0.61 & 0.43 & 0.21 & 0.15 \\ 
		Heroin & 152.79 & 0.65 & 0.51 & 0.28 & 0.22 \\ 
		LSD & 5.69 & 0.77 & 0.71 & 0.56 & 0.49 \\ 
		Ketamine & 79.36 & 0.58 & 0.54 & 0.42 & 0.37 \\ 
		\bottomrule
	\end{tabular*}
	\fnote{\textbf{Notes:} The table reports the discount rates for the eight product categories by quantity. Prices are reported in USD. The bitcoin exchange rate used corresponds to the day on which the item price was observed. Prices reported are the unit price, that is, the price per consumption unit, defined as 1 gram for all categories except LSD, where it is 10 $\times$ 100 $\mu$g.}
\end{table}

\begin{table}
	\captionsetup{width=\textwidth,justification=raggedright,singlelinecheck=off}
	\caption{Linked vendor accounts}
	\label{table:sellers}
	\small
	\begin{tabular*}{\textwidth}{@{\extracolsep{2.5pt}} lcc@{~}cccc}
		\toprule
		\addlinespace[0.5em]
		& & \multicolumn{5}{c}{Identified sellers} \\
		\cmidrule(rl){3-7}
		& accounts & total & one account & two accounts & three accounts & four accounts \\
		\hline \\[-1.8ex]
		N & 3,005 & 2,344 & 1,718 & 620 & 23 & 3\\
		\bottomrule
	\end{tabular*}
	\fnote{\textbf{Notes:} The table shows the number of vendor accounts and the number of identified sellers present in total and by the number of accounts sellers use on the two platforms. There are no sellers active on only one platform with multiple accounts.}
\end{table}

 \begin{table}
   \centering
	\captionsetup{width=\textwidth,justification=raggedright,singlelinecheck=off}
	\caption{Sellers on Evolution before the exit}
	\label{table:sw-pre-exit}
	\small
	\begin{tabular*}{\textwidth}{@{\extracolsep{1pt}} lccccrrcc}
		\toprule
		\addlinespace[1em]
		& \multicolumn{2}{c}{Share of sellers} & \multicolumn{2}{c}{Avg. \# offers} & \multicolumn{2}{c}{Median quantity} & \multicolumn{2}{c}{Average Rating}\\
		\cmidrule{2-3} \cmidrule{4-5} \cmidrule{6-7} \cmidrule{8-9}
		Category & Re-entr.\ & Evol.\ & \multicolumn{1}{c}{Re-entr.} & \multicolumn{1}{c}{Evol.} & Re-entr.\ & Evol.\ & Re-entr.\ & Evol.\ \\
		\hline \\[-1.8ex]
		Cocaine & $0.35$ & $0.27$ & $2.68$ & $3.32$ & $2$ & $3$ & $99.49$ & $97.43$ \\ 
		Cannabis & $0.67$ & $0.69$ & $2.35$ & $3.41$ & $7.09$ & $10$ & $99.66$ & $97.64$ \\ 
		Amphetamine & $0.15$ & $0.20$ & $2.50$ & $3.84$ & $25$ & $20$ & $98.91$ & $98.71$ \\ 
		\bottomrule
	\end{tabular*}
	\fnote{\textbf{Notes:} The table contrasts re-entering sellers to all other Evolution sellers in the month prior to the exit, excluding the last seven days. It shows the share of the two groups in each product category, the mean number of offers, the median quantity sold, and the mean rating.}
\end{table}

%



\begin{table}
	\centering
	\captionsetup{width=\textwidth,justification=raggedright,singlelinecheck=off}
	\caption{Sensitivity analysis: The effect of re-entry on rating, specification without controls}
	\small
	\begin{threeparttable}
		\begin{tabular*}{\textwidth}{@{\extracolsep{\fill}} l*{5}{S[%
					table-number-alignment=center,
					table-figures-integer=2, table-figures-decimal=3,
					round-precision=3,
					table-space-text-pre={-},
					table-space-text-post={***}
					]}}
			\toprule
			\addlinespace[2em]
			& {All} & {Cannabis} & {Cocaine} & {Amphetamine} \\
			\midrule
			Re-entry x post         & -0.087{**} & -0.091  & -0.089  & -0.077  \\
									& (0.041)     & (0.065) & (0.073) & (0.077) \\
									& [0.035]	 & [0.161] & [0.221] & [0.314] \\
			\midrule
			Item-specific FE        & {\checkmark} & {\checkmark} & {\checkmark} & {\checkmark} \\	
			Time FE        			& {\checkmark} & {\checkmark} & {\checkmark} & {\checkmark} \\		
			Num. obs.               & {50254}	   & {27220}   & {14947}   & {8087}    \\
			Num. items\    			& {6656}        & {3708}    & {1854}    & {1094}    \\
			\bottomrule
		\end{tabular*}
		\begin{tablenotes}[para,flushleft]\scriptsize
		  \item \textbf{Notes:} Results based on a linear model as specified in \autoref{sec:Approach} with weekly data. The sample is restricted to a time period around the evolution exit, from 15 weeks prior to 10 weeks after the exit. Standard errors clustered at seller-substance level given in parentheses, $p$-values in brackets. *, ** and *** denote p$<$0.1, p$<$0.05 and p$<$0.01, respectively.
		\end{tablenotes}
	\end{threeparttable}
	\label{tab:fs-nc}
\end{table}


\begin{table}
\centering
\captionsetup{width=\textwidth,justification=raggedright,singlelinecheck=off}
\caption{Sensitivity analysis: The effect of rating on price, specification without controls}
\small
\begin{threeparttable}
	\begin{tabular*}{\textwidth}{@{\extracolsep{\fill}} l*{5}{S[%
	table-number-alignment=center,
	table-figures-integer=2, table-figures-decimal=3,
	round-precision=3,
	table-space-text-pre={-},
	table-space-text-post={***}
	]}}
	\toprule
	\addlinespace[2em]
	 & {All} & {Cannabis} & {Cocaine} & {Amphetamine} \\
	\midrule
	Rating (ihs)            & 1.036{*} & 0.750   & 0.484   & 2.257   \\
							& (0.551)   & (0.593) & (0.566) & (2.376) \\
							& [0.060]	& [0.206]  & [0.392]  & [0.342] \\						
	\midrule
	Item-specific FE        & {\checkmark} & {\checkmark} & {\checkmark} & {\checkmark} \\	
	Time FE        			& {\checkmark} & {\checkmark} & {\checkmark} & {\checkmark} \\		
	Num. obs.               & {50254}     & {27220}   & {14947}   & {8087}    \\
	Num. items     & {6656}      & {3708}    & {1854}    & {1094}    \\
	\bottomrule
	\end{tabular*}
	\begin{tablenotes}[para,flushleft]\scriptsize
	  \item \textbf{Notes:} Results based on a linear model as specified in \autoref{sec:Approach} with weekly data. The sample is restricted to a time period around the evolution exit, from 15 weeks prior to 10 weeks after the exit. Standard errors clustered at item-level given in parentheses, $p$-values in brackets. *, ** and *** denote p$<$0.1, p$<$0.05 and p$<$0.01, respectively.
	\end{tablenotes}
\end{threeparttable}
\label{tab:iv-nc}
\end{table}


\begin{table}
	\centering
	\captionsetup{width=\textwidth,justification=raggedright,singlelinecheck=off}
	\caption{Sensitivity analysis: The effect of rating on price (OLS)}
	\small
	\begin{threeparttable}
		\begin{tabular*}{\textwidth}{@{\extracolsep{\fill}} l*{5}{S[%
					table-number-alignment=center,
					table-figures-integer=2, table-figures-decimal=3,
					round-precision=3,
					table-space-text-pre={-},
					table-space-text-post={***}
					]}}
			\toprule
			\addlinespace[2em]
			& {All} & {Cannabis} & {Cocaine} & {Amphetamine} \\
			\midrule			
			Rating (ihs)            & -0.006    & 0.013{**} & -0.052     & -0.009  \\
									& (0.010)   & (0.006)    & (0.049)    & (0.022) \\
									& [0.519]	& [0.043]	& [0.289]	& [0.671]	\\
			\midrule
			Item-specific FE        & {\checkmark} & {\checkmark} & {\checkmark} & {\checkmark} \\	
			Time FE        			& {\checkmark} & {\checkmark} & {\checkmark} & {\checkmark} \\		
			Num. obs.               & {50254}     & {27220}     & {14947}      & {8087}    \\
			Num. gitem      & {6656}      & {3708}      & {1854}       & {1094}    \\
			\bottomrule
		\end{tabular*}
		\begin{tablenotes}[para,flushleft]\scriptsize
		  \item \textbf{Notes:} Results based on a linear model as specified in \autoref{sec:Approach} with weekly data. The sample is restricted to a time period around the evolution exit, from 15 weeks prior to 10 weeks after the exit. Standard errors clustered at seller-substance level given in parentheses, $p$-values in brackets. *, ** and *** denote p$<$0.1, p$<$0.05 and p$<$0.01, respectively.
		\end{tablenotes}
	\end{threeparttable}
	\label{tab:ols}
\end{table}


\begin{table}
	\centering
	\captionsetup{width=\textwidth,justification=raggedright,singlelinecheck=off}
	\caption{Sensitivity analysis: The effect of rating on price with a placebo exit}
	\small
	\begin{threeparttable}
		\begin{tabular*}{\textwidth}{@{\extracolsep{\fill}} l*{5}{S[%
					table-number-alignment=center,
					table-figures-integer=2, table-figures-decimal=3,
					round-precision=3,
					table-space-text-pre={-},
					table-space-text-post={***}
					]}}
			\toprule
			\addlinespace[2em]
			& {All} & {Cannabis} & {Cocaine} & {Amphetamine} \\
			\midrule		
			Rating (ihs)            & -9.366   & -1.862   & 13.778   & -56.175   \\
									& (13.518) & (12.383) & (23.578) & (192.288) \\
									& [0.488]  & [0.880]  & [0.559]  & [0.770]	\\
			\midrule
			Item-specific FE        & {\checkmark} & {\checkmark} & {\checkmark} & {\checkmark} \\	
			Time FE        			& {\checkmark} & {\checkmark} & {\checkmark} & {\checkmark} \\		
			Num. obs.               & {56487}    & {30661}    & {16076}    & {9750}      \\
			Num. items      & {6836}     & {3860}     & {1916}     & {1060}      \\
			\bottomrule
		\end{tabular*}
		\begin{tablenotes}[para,flushleft]\scriptsize
		  \item \textbf{Notes:} Results based on a linear model as specified in \autoref{sec:Approach} with weekly data. The sample is restricted to a time period that ends a week before the exit. We assume a placebo exit occurs 10 weeks prior and consider the 15 weeks leading up to the placebo exit and 10 weeks after. Standard errors clustered at seller-substance level given in parentheses, $p$-values in brackets. *, ** and *** denote p$<$0.1, p$<$0.05 and p$<$0.01, respectively.
		\end{tablenotes}
	\end{threeparttable}
	\label{tab:pl}
\end{table}






\clearpage

\onehalfspacing
\bibliographystyle{custom_natbib}
\bibliography{dn}

\end{document}